\documentclass[10pt,paper]{article}
\usepackage[pdftex]{graphicx}
\usepackage{amsthm,amsfonts,amsmath,amssymb}
\usepackage{caption}

\begin{document}


\begin{center}
{\bf {\Large Description of 4 Spacecraft, Moving}}

{\bf {\Large on Elliptic Kepler Orbits}}


\end{center}

\noindent
\begin{center}
Vladimir P. Zhukov$^{a,b,*}$, Nikolai K. Iakovlev$^{a,c}$, Alexander A. Bochkarev$^{c}$,

Nikita E. Logvinenko$^{c}$, Sergei M. Kurchev$^{c}$, Vlas A. Karavaikin$^{c}$, Ivan A. Radko$^{c}$
 \end{center}

%
%

\noindent {}

\noindent${}^{a}$\textit{ Federal Research Center for Information and Computational Technologies, Novosibirsk, 630090, Russia}

\noindent ${}^{b}$\textit{ Novosibirsk State Technical University, Novosibirsk, 630073, Russia}

\noindent ${}^{c}$\textit{ Novosibirsk State University, Novosibirsk, 630090, Russia}

\begin{center}
$^*$Corresponding author: \textit{$zukov@ict.nsc.ru$.}
\end{center}

\noindent {}

\noindent {\small {\bf Abstract.}
The four-spacecraft formation is essential for measurements of various physical fields. The use of this formation on substantially elliptical heliocentric Kepler orbits allows measuring gradients of gravitation field in Solar system. The accuracy of the measurements will be sufficient to confirm or to refute modified theories of gravity. In this paper a new approach for the description of this formation is presented. The analytical solutions of the linearized motion equations are obtained. The distinctive feature of the solutions is that they use Cartesian coordinates of one of the spacecraft, termed the chief. These solutions have a clear physical meaning. It is shown, that the volume of a tetrahedron formed by spacecraft is a polynomial of 3-rd degree of Cartesian coordinates of the chief. The polynomial's coefficients are functions of initial spacecraft coordinates and velocities and linearly depend on time. If all spacecraft have the same periods of rotation around the Sun, the volume is a polynomial of 2-nd degree of the chief coordinates with time-independent coefficients.  In this case the volume can be zeroed from 0 to 4 times per the period. Suggested approach can significantly simplify planning missions for measurements of various interplanetary fields.}

\noindent {}

\noindent \textit{Keywords: }Tetrahedral formation; Elliptical orbits; Linear approximation; Tetrahedron volume; Tetrahedron quality

\noindent {}

\begin{center} \textbf{1. Introduction} \end{center}

\noindent Measurements of the spatiotemporal distributions of physical fields in the circum- and interplanetary space require at least four spacecraft [1]. Tetrahedral formations of spacecraft, moving around the Earth are intensively studied and used in practice to study electromagnetic fields, plasma density and other quantities in Earth's magnetosphere [1-12].

In [13-14] it is proposed to use a tetrahedral spacecraft formation rotating around the Sun on substantially elliptical orbits to measure gravity field gradients. The accuracy of the measurements might be sufficient to detect dark matter and test modified gravity theories, such as the Yukawa theory [15] and the Galileon theory [16-19]. The requirements for the formation are as follows:

1. To save fuel, the spacecraft must move freely, i.e. the orbits must be Keplerian.

2. The orbits must be substantially elliptical, not circular. This is necessary in order to measure the gravitational force at different distances from the Sun. An eccentricity close to 1 would result in the spacecraft coming too close to the Sun at perihelion. This is not acceptable from a thermal standpoint. The assumed eccentricity of the orbits is 0.6.

3. The semi-major axis of the orbit must be equal to 1 astronomical unit (AU) or about 150 million kilometers for reasons related to the thermal regime of spacecraft and the technical feasibility of the mission.

4. All spacecraft should have equal periods of rotation around the Sun to repeat measurements in the case of failure on the first circuit.

To perform optical measurements, it is necessary that

5. The distances between the spacecraft are about 1000 km.

6. The volume of the tetrahedron with vertices at the location of the spacecraft (hereinafter referred to as ``tetrahedron formation'', ``tetrahedron'') should not become close to zero.

7. Ideally, the shape of the tetrahedron would be close to regular. However, the issue of restrictions on the shape requires further study.

To plan specific missions, it is necessary to have an idea of the evolution of the formation and have a convenient mathematical tool for solving various optimization problems. To create the tool, one of the spacecraft is selected as the chief, and its orbit is called the reference orbit. Any of the four spacecraft can be the chief. For convenience, the chief spacecraft numbered as 0. The remaining 3 spacecraft are called deputies and numbered $m = 1, 2, 3$. The motion of the deputy is described by relative coordinates and velocities -- the differences between their coordinates and velocities and those of the chief.

Since the distances between the spacecraft are much smaller than the size of the orbits, linearization of the problem is natural. A small parameter -- the ratio of the relative coordinates to the characteristic size of the reference orbit -- is used.  The ratio of the relative velocities to the characteristic velocity of the chief also assumed to be proportional to this parameter. This approach is called a linear approximation, and the corresponding equations of motion are called linearized or linear.

Solving linear equations of motion, especially for the evolution of several spacecraft relative positions, is complex. For two spacecraft, this problem can be solved using Keplerian orbit elements as coordinates [20-24] (Tschauner-Hempel equations [20]). The resulting formulas are difficult to analyze. For example, in [13] these formulas are given, but the linearized equations of motion are solved using the Runge-Kutta method when performing specific calculations.

An important quantity for measuring various physical fields is the volume of the tetrahedron, formed by the spacecraft. It is desirable that this quantity does not vanish over as large part of the orbit as possible. The volume evolution in the case of orbits close to circular is studied in detail in [2,3].  In particular it is shown, that formations with non-vanishing volume exist for such orbits. The case of essentially elliptical orbits has not been systematically studied. The existence of formations with non-vanishing volume for elliptical orbits was not known.

This paper presents a new mathematical tool for studying in a linear approximation the tetrahedral formations of spacecraft, moving along essentially elliptical orbits. The uniqueness of the tool is in its use of Cartesian coordinates fixed relative to the Sun (or another object taken as a fixed gravitational center). Moreover, the relative coordinates and velocities at different points of the orbit are expressed not through time, but through the Cartesian coordinates of the chief. Three deputy spacecraft are considered as a single object (vector). This approach allows identifying symmetries in the evolution of the formation. It has a number of advantages (see the Discussion section).

It is shown for the first time that in the case of essentially elliptical orbits the volume of a tetrahedron is a third-order polynomial of the Cartesian coordinates of the chief. The coefficients of this polynomial are functions of the initial relative coordinates and velocities of the deputies. The coefficients depend linearly on time also.  When the spacecraft have equal periods of revolution around the Sun, the volume of the tetrahedron is a second-degree polynomial with time-independent coefficients. The volume can vanish from zero to four times per period.

Further, the work describes the system of used coordinates, designations, and dimensionless variables. The dependencies of the coordinates and velocities of the chief on time are given, as well as important relationships between them, which are used further to derive formulas. Then, linearized equations for the relative coordinates of the deputies and solutions to these equations are presented. The solutions are obtained from considerations of various types of perturbations of the reference orbit. Expression for the volume of the tetrahedron is presented. Next, an important case of equal rotation periods of all spacecraft is considered. All formulas are significantly simplified if the moment when the chief is at perihelion is chosen as the initial moment of time. Then, the developed methodology is used to provide examples of studying the evolution of tetrahedral spacecraft formation.  The Discussion describes the limits of applicability, advantages of the developed approach, as well as possible features of its application for planning and optimization of missions, and increasing the accuracy. Finally, the Conclusion is presented.

\begin{center} \textbf{2. Methodology} \end{center}

\begin{center} \textit{2.1. Coordinate System, Designations, Normalization} \end{center}

\begin{figure}
\centering {\includegraphics*[width=7cm]{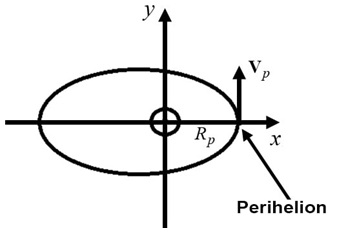}}
\caption*{\textbf{Fig. 1. }Coordinate system. $R_{p} $ and $V_{p} $ are distance to the Sun and velocity of the chief at perihelion.}
\end{figure}


To describe the spacecraft positions, a Cartesian coordinate system centered on the Sun is used. The axes $x$, $y$ are placed in the plane of the reference orbit, $z$ -- across it. The axis $x$ is directed from the center of coordinates to the perihelion of the reference orbit. A right-hand coordinate system is used. It is assumed that the chief rotates counterclockwise. The following notations will be used (see Appendix A):

\noindent ${\bf R}=[X,Y,Z]$ is a vector of the chief position. $Z=0$, $R=\sqrt{X^2 +Y^2} $.

\noindent ${\bf V}=\dot{{\bf R}}=[\dot{X},\dot{Y},\dot{Z}]$ is a velocity of the chief, $\dot{Z}=0$.

\noindent ${\bf R}_{m} $ is a vector of the position of the deputy with number $m$, $m=1,2,3$.

\noindent $[x_m,y_m,z_m]\equiv {\bf r}_m \equiv {\bf R}_m-{\bf R}$ are relative coordinates of the deputy with number $m$.

\noindent $u_{mx} ,u_{my} ,u_{mz} $ are components of the relative velocity ${\bf u}_m =\dot{{\bf r}}_m $ of the deputy with number $m$.

\noindent $G=\gamma M$ is a factor in the law of gravity, $\gamma $ and $M$ are the gravitational constant and the mass of the Sun respectively.

The equation of motion of a spacecraft in the gravitational field of the Sun has the form $\ddot{{\bf R}}=-G{\bf R}/R^{3} $. Let the semi-major axis of the reference orbit\textit{ a} be the length scale, the time $\bar{t}=2\pi t/T$ be the time scale (the dimensionless period is equal to $2\pi $), the speed  $2\pi a/T$ be the velocity scale. Here $T=2\pi \sqrt{a^{3} /G} $ is the period of revolution of the chief about the Sun. Then, the equation of motion in dimensionless form is $d^{2} \bar{{\bf R}}/d\bar{t}^{2} =-\bar{{\bf R}}/\bar{R}^{3} $. In what follows, unless otherwise noted, dimensionless variables will be used everywhere and the bar will be omitted.

\noindent

\begin{center} \textit{2.2. Motion of the chief } \end{center}

\noindent It is necessary to solve the following problem to calculate the dimensionless coordinates of the chief, which is at perihelion at time $t=0$:
$$Z=0,$$
\begin{equation} \label{GrindEQ__1_}
\ddot{X}=-X(X^2 +Y^2)^{-3/2} ,\ \ t=0{\text :} \ X=R_p, \ \dot{X}=0,
\end{equation}
\begin{equation} \label{GrindEQ__2_}
\ddot{Y}=-Y(X^{2} +Y^{2} )^{-3/2} ,\ \ t=0{\text :} \ Y=0, \ \dot{Y}=V_{p},
\end{equation}
\begin{equation} \label{GrindEQ__3_}
R_{p} =(1-e),\ \ V_{p} =\sqrt{\frac{1+e}{1-e} } ,
\end{equation}
where$R_p$ and $V_p$ are the radius and speed of the spacecraft in perihelion, $e$ is the eccentricity.

The solution to this problem has the form [25]:
\begin{equation} \label{GrindEQ__4_}
X=\cos \xi -e,\ \ Y=\sqrt{1-e^2} \sin \xi  ,
\end{equation}
\begin{equation} \label{GrindEQ__5_}
\xi -e\sin \xi =t,
\end{equation}
where $\xi $ is a parameter.

Formulas \eqref{GrindEQ__4_} and \eqref{GrindEQ__5_} allow deriving useful relationships, including expressions for time-derivatives of the coordinates in terms of the coordinates themselves:
\begin{equation} \label{GrindEQ__6_}
R=1-e(X+e),   X+e=\frac{1-R}{e}  ,
\end{equation}
\begin{equation} \label{GrindEQ__7_}
\dot{\xi }=(1-e\cos \xi )^{-1} =\, R^{-1} ,
\end{equation}
\begin{equation} \label{GrindEQ__8_}
\dot{X}=-\frac{Y}{R\sqrt{1-e^{2} } }  ,
\end{equation}
\begin{equation} \label{GrindEQ__9_}
\dot{Y}=\frac{(X+e)}{R} \sqrt{1-e^2} =\frac{1-R}{eR} \sqrt{1-e^2} ,
\end{equation}
\begin{equation} \label{GrindEQ__10_}
\dot{R}=-e\dot{X}=\frac{eY}{R\sqrt{1-e^2} } ,
\end{equation}
\begin{equation} \label{GrindEQ__11_}
V^{2} =(\dot{X})^{2} +(\dot{Y})^{2} =(2-R)/R,
\end{equation}
\begin{equation} \label{GrindEQ__12_}
X\dot{X}+Y\dot{Y}=eY(1-e^{2} )^{-1/2}  ,
\end{equation}
\begin{equation} \label{13}
(X+e)^{2} +Y^{2} /(1-e^{2} )=1 {\mbox {\text { (orbit equation)}}},
\end{equation}
\begin{equation} \label{14}
V^{2} /2-1/R=-1/2 {\mbox {\text { (the energy conservation law)}}},
\end{equation}
\begin{equation} \label{15}
X\dot{Y}-Y\dot{X}=\sqrt{1-e^{2} }  {\mbox {\text { (the law of conservation of angular momentum)}}}.
\end{equation}

\noindent {}

\noindent {}

\begin{center} \textit{2.3. The Linearized Problem for Determining the Coordinates of Deputies}\end{center}

\noindent Since the distances between the spacecraft (about 1,000 km) are much smaller than the size of the reference orbit (150 million km), it is reasonable to linearize the spacecraft equations of motion relative to the chief. Assuming $r_m \ll R$ and $u_m \ll V$, the equations of motion $\ddot{R}_m =-R_m /R_m^{3} $ in the linear approximation can be transformed into
\begin{equation} \label{GrindEQ__16_}
\ddot{x}_m =-\frac{1}{R^{3} } \left(1-3\frac{X^{2} }{R^{2} } \right)x_m +3\frac{XY}{R^{5} } y_m ,
\end{equation}
\begin{equation} \label{GrindEQ__17_}
\ddot{y}_m =-\frac{1}{R^{3} } \left(1-3\frac{Y^{2} }{R^{2} } \right)y_m +3\frac{XY}{R^{5} } x_m ,
\end{equation}
\begin{equation} \label{GrindEQ__18_}
\ddot{z}_m =-z_m /R^{3} ,
\end{equation}
$$[u_{mx} ,u_{my} ,u_{mz} ]=[\dot{x}_m ,\dot{y}_m ,\dot{z}_m ]. $$
Thus, the equation for $z_{m} $ is separated from the equations for $x_m ,y_m$. Equations (16-18) are supplemented by initial conditions according to which, at a given moment of time $t_0 $ the quantities $x_m,y_m,z_m$, $u_{mx},u_{my},u_{mz} $ are specified.

The applicability of linear equations (16-18), in addition to the natural condition of smallness of the relative distances and velocities compared to the size of the reference orbit and the speed of the chief, has extra limitations:

1. Linearity will be violated after many revolutions in the case of orbits with different periods. However, this case is not interesting, since orbits with the same periods and a small number of revolutions are of the greatest interest.

2. If the formation is such that, for example, the differences of the angle of a triangle formed by 3 spacecraft is close to $\pi $ or 0 (degeneracy), then the evolution of this formation and, in particular, the volume of the tetrahedron, is determined by smallness of 2-nd order in $r_{m} /R$. Such a configuration requires a higher order approximation for its calculation, but it is also not interesting from the point of planning the missions.

\noindent {}

\begin{center} \textit{2.4. Solution to the Linearized Equations} \end{center}

\noindent {}

To solve Eq.-s \eqref{GrindEQ__16_}-\eqref{GrindEQ__18_} we note that problem \eqref{GrindEQ__1_}, \eqref{GrindEQ__2_} can be written as
\begin{equation} \label{GrindEQ__19_}
\ddot{X}=-X/R^{3} ,\ t=0{\text :}\ \ X=R_p ,\ \ \dot{X}=0,
\end{equation}
\begin{equation} \label{GrindEQ__20_}
\ddot{Y}=-Y/R^{3} ,\ \ t=0{\text :}\ \ Y=0,\ \ \dot{Y}=V_{p} .
\end{equation}
Assuming $R=\sqrt{X^{2} +Y^{2} } $ is a given function of time $R(t)$, and comparing \eqref{GrindEQ__19_}, \eqref{GrindEQ__20_} with \eqref{GrindEQ__18_}, it can be seen that the general solution \eqref{GrindEQ__18_} takes the form
\begin{equation} \label{GrindEQ__21_}
z_{m} =\alpha _m X+\beta _m Y
\end{equation}
and one of the particular solutions of \eqref{GrindEQ__16_}, \eqref{GrindEQ__17_} is
\begin{equation} \label{GrindEQ__22_}
\left[\begin{array}{l} {x_m } \\ {y_m } \end{array}\right]={\bf r}_{\chi } \chi _{m} ,{\rm \; \; }{\bf r}_{\chi } =\left[\begin{array}{l} {x_{\chi } } \\ {y_{\chi } } \end{array}\right]=\left[\begin{array}{l} {-Y} \\ {X} \end{array}\right], \chi _m ={\rm const}.
\end{equation}
Solutions \eqref{GrindEQ__21_}, \eqref{GrindEQ__22_} can be interpreted as rotations of the reference orbit around different axes by small angles. Adding to them small deformations associated with a time delay, a change in eccentricity and a change in the major semi-axis (or, what is the same, the energy or period of revolution), a fundamental system of solutions of equations (16-18) is obtained.

1.  Rotation around the $y$-axis by a small angle $\alpha _m$, taking into account that, for the reference orbit $Z=0$, yields the following:
$$\left[\begin{array}{l} {x_m } \\ {z_m } \end{array}\right]=\left[\begin{array}{cc} {\cos \alpha _m} & {-\sin \alpha _m} \\ {\sin \alpha _m} & {\cos \alpha _m} \end{array}\right]\left[\begin{array}{l} {X} \\ {Z} \end{array}\right]-\left[\begin{array}{l} {X} \\ {Z} \end{array}\right]\approx \left[\begin{array}{cc} {0} & {-\alpha _m } \\ {\alpha _m} & {0} \end{array}\right]\left[\begin{array}{c} {X} \\ {0} \end{array}\right]=$$
$$=\left[\begin{array}{c} {0} \\ {X} \end{array}\right]\alpha _m .$$
2.  Similarly, rotation around the $x$-axis by a small angle $\beta _m$ gives $z_m=Y\beta _m$.

\noindent 3.  The change in coordinates $x$, $y$ due to rotation the reference orbit around the $z$-axis by an angle $\chi _m$ is given by formula \eqref{GrindEQ__22_}:
\begin{equation*}
\left[\begin{array}{l} {x_m} \\ {y_m} \end{array}\right]=\left[\begin{array}{cc} {\cos \chi _m} & {-\sin \chi _m} \\ {\sin \chi _m} & {\cos \chi _m} \end{array}\right]\left[\begin{array}{l} {X} \\ {Y} \end{array}\right]-\left[\begin{array}{l} {X} \\ {Y} \end{array}\right]\approx \left[\begin{array}{cc} {0} & {-\chi _m} \\ {\chi _m } & {0} \end{array}\right]\left[\begin{array}{l} {X} \\ {Y} \end{array}\right]={\bf r}_{\chi } \chi _m .
\end{equation*}
4.   Time shift by $\tau _{m} $
$$\left[\begin{array}{l} {x_m (t)} \\ {y_m (t)} \end{array}\right]=\left[\begin{array}{c} {X(t+\tau _m )} \\ {Y(t+\tau _m )} \end{array}\right]-\left[\begin{array}{c} {X(t)} \\ {Y(t)} \end{array}\right]\approx \left[\begin{array}{l} {\dot{X}} \\ {\dot{Y}} \end{array}\right]\tau _m . $$
Accordingly,
\begin{equation} \label{GrindEQ__23_}
{\bf r}_{\tau } =\left[\begin{array}{l} {x_{\tau } } \\ {y_{\tau } } \end{array}\right]=\left[\begin{array}{l} {\dot{X}} \\ {\dot{Y}} \end{array}\right].
\end{equation}
5.   Change in eccentricity $e\to e+\eta $ with a constant semi-axis $a$ (i.e. constant energy and period of revolution around the Sun) and with the preserved orientation of the semi-axes gives (see Appendix B.)
\begin{equation} \label{GrindEQ__24_}
\left[\begin{array}{l} {x_m} \\ {y_m } \end{array}\right]={\bf r}_{\eta } \eta _m,\ \ {\bf r}_{\eta } =\left[\begin{array}{l} {x_{\eta } } \\ {y_{\eta } } \end{array}\right]=\left[\begin{array}{l} {\partial X/\partial e} \\ {\partial Y/\partial e} \end{array}\right]=\left[\begin{array}{c} {-\left(1+\frac{Y^{2} }{R(1-e^{2} )} \right)} \\ {\frac{XY}{R(1-e^{2} )} } \end{array}\right]
\end{equation}
The function $1+Y^{2} \left(R(1-e^{2} )\right)^{-1} $ is strictly greater than 0. That is, the difference in coordinates \textit{x} of the spacecraft whose orbits differ only in eccentricity, has a constant sign and is not less than this difference in perihelion and aphelion. For coordinates $y$ and $z$, as well as for coordinate $x$, but with other types of changes in the reference orbit, the coordinates difference changes sign when moving along the orbit (see (21-25)).

\noindent 6.  Change in the semi-major axis $a$ (energy, period of revolution) at constant eccentricity. The orbit increases or decreases, preserving the shape and orientation of the semi-axes. In this case, it is necessary to return to dimensional quantities. In Appendix C shown that for this case
\begin{equation} \label{GrindEQ__25_}
\left[\begin{array}{l} {x_m} \\ {y_m} \end{array}\right]={\bf r}_{\upsilon } \upsilon _m ,\ \ {\bf r}_{\upsilon } =\left[\begin{array}{l} {x_{\upsilon } } \\ {y_{\upsilon } } \end{array}\right]=\left[\begin{array}{l} {\partial X/\partial a} \\ {\partial Y/\partial a} \end{array}\right]=\left[\begin{array}{c} {X-(3/2)t\dot{X}} \\ {Y-(3/2)t\dot{Y}} \end{array}\right]
\end{equation}
Solution \eqref{GrindEQ__25_} contains time as a multiplier. This is a result of the difference in the period of revolution of the deputies from the period of the chief.

Direct substitution of (21-25) into the equations (16-18) and comparison with the solutions obtained by finite-difference methods confirm the correctness of these results.

Functions ${\bf r}_\chi ,{\bf r}_\tau ,{\bf r}_\eta ,{\bf r}_\upsilon $ form a fundamental system of solutions to equations \eqref{GrindEQ__16_}, \eqref{GrindEQ__17_}, and the functions $X,Y$ form a fundamental system of solutions to equation \eqref{GrindEQ__18_}. The coefficients $\alpha _m ,\beta _m ,\chi _m ,\tau _m ,\eta _m ,\upsilon _m$ are constants of integration of the corresponding equations.

Thus, the solution of the linear problem for coordinates in the plane of the reference orbit can be expressed in dimensionless variables as
$$\left[\begin{array}{c} {x_m} \\ {y_m} \\ {u_{xm} } \\ {u_{ym} } \end{array}\right]=A\left[\begin{array}{c} {\chi _m } \\ {\tau _m} \\ {\eta _m} \\ {\upsilon _m } \end{array}\right], \ \  A=\left[\begin{array}{cccc} {x_\chi } & {x_\tau } & {x_\eta } & {x_\upsilon } \\ {y_\chi } & {y_\tau } & {y_\eta } & {y_\upsilon } \\ {\dot{x}_\chi } & {\dot{x}_{\tau } } & {\dot{x}_{\eta } } & {\dot{x}_{\upsilon } } \\ {\dot{y}_{\chi } } & {\dot{y}_{\chi } } & {\dot{y}_{\eta } } & {\dot{y}_{\upsilon } } \end{array}\right]$$
or
\begin{equation} \label{GrindEQ__26_}
A=\left[\begin{array}{cccc} {-Y} & {\dot{X}} & {-\left(1+\frac{Y^{2} }{R(1-e^{2} )} \right)} & {X-(3/2)t\dot{X}} \\ {X} & {\dot{Y}} & {\frac{XY}{R(1-e^{2} )} } & {Y-(3/2)t\dot{Y}} \\ {-\dot{Y}} & {-\frac{X}{R^{3} } } & {\frac{(X+e+X/R)\dot{X}}{R} } & {-(1/2)\dot{X}+\frac{(3/2)Xt}{R^{3} } } \\ {\dot{X}} & {-\frac{Y}{R^{3} } } & {\frac{X\dot{Y}}{R(1-e^{2} )} +\frac{Y\dot{X}}{R^{2} } } & {-(1/2)\dot{Y}+\frac{(3/2)Yt}{R^{3} } } \end{array}\right]
\end{equation}
Here formulas (19), (20), \eqref{GrindEQ__8_}, \eqref{GrindEQ__9_}, (10), (12), (22-25) were used.

It can be shown that for $e\ne 0$, the matrix $A$ is invertible at any time. For \textit{e}=0, the determinant $A$ is zero, because rotation in the orbital plane and the time shift coincide for circular orbits: ${\bf r}_{\chi } ={\bf r}_{\tau } $. In this case, instead of ${\bf r}_{\chi } $ and ${\bf r}_{\tau } $, the functions ${\bf r}_{\chi } $ and, for example,
$${\bf r}_\nu =\mathop{\lim }\limits_{e\to 0} \frac{{\bf r}_\tau -{\bf r}_\chi }{e} =\left[\begin{array}{c} {-XY} \\ {X^{2} +1} \end{array}\right]$$
can be taken.

\noindent Solution ${\bf r}_\nu $ corresponds to a change in the eccentricity of a circular orbit along the axis $y$, solution ${\bf r}_\eta $ --  along the axis $x$. The solution ${\bf r}_\nu \equiv ({\bf r}_\tau -{\bf r}_\chi )/e$ can also be used for nonzero $e$. However, it is more cumbersome and less clear than ${\bf r}_\tau  $ in this case.

The coefficients $\alpha _m$, $\beta _m$, $\chi _m$, $\tau _m$, $\eta _m$, $\upsilon _m$ are found from the initial conditions. Setting $x_m ,y_m ,z_m$, $u_{xm} ,u_{ym} ,u_{zm} $ at the initial time $t=t_0$, and using \eqref{GrindEQ__21_} and \eqref{GrindEQ__26_} we find $(\alpha _m$, $\beta _m$, $\chi _m$, $\tau _m$, $\eta _m$, $\upsilon _m)$. With $(\alpha _m ,\beta _m ,\chi _m ,\tau _m ,\eta _m ,\upsilon _m )$ obtained and using \eqref{GrindEQ__21_} and \eqref{GrindEQ__26_} we can find $(x_m ,y_m ,z_m ,u_{xm} ,u_{ym} ,u_{zm} )$ at an arbitrary time.

Particularly, the initial conditions $z_m =z_m (t_0)$, $\dot{z}_m =u_{zm} (t_0)$, at the time $t=t_0$ give:
$$z_m (t)=\frac{\left(\left. \dot{Y}\right|_{t=t_0} X(t)-\left. \dot{X}\right|_{t=t_{0} } Y(t)\right)z_m (t_0)-\left(Y(t_0)X(t)-X(t_0)Y(t)\right)u_{zm} (t_0)}{\left. \dot{Y}\right|_{t=t_0} X(t_0)-\left. \dot{X}\right|_{t=t_0} Y(t_0)}$$
or, taking into account (15):
\begin{equation} \label{GrindEQ__27_}
\alpha _m =\frac{\left. \dot{Y}\right|_{t=t_0} z_{m} (t_0)-Y(t_0)u_{zm} (t_0)}{\sqrt{1-e^{2} } } ,  \beta _m =\frac{-\left. \dot{X}\right|_{t=t_{0} } z_m (t_0)+X(t_0)u_{zm} (t_0)}{\sqrt{1-e^{2} } }
\end{equation}

\noindent
\begin{center} \textit{2.5. Evolution of the Volume of the Tetrahedron in the Linear Approximation}\end{center}

The volume V of a tetrahedron is calculated according to the formula
$${\rm V}=|D|/6, \ D=\det \left|\begin{array}{ccc} {x_1} & {y_{1} } & {z_1} \\ {x_2} & {y_2} & {z_2} \\ {x_3} & {y_3} & {z_3} \end{array}\right|\equiv \det \left|\begin{array}{ccc} {x_1} & {x_2} & {x_{3} } \\ {y_1} & {y_2} & {y_3} \\ {z_1} & {z_2} & {z_3} \end{array}\right|$$
The last expression can be interpreted as a determinant of matrix, the rows of  which are the vectors ${\bf x}=(x_1,x_2,x_3)$, ${\bf y}=(y_1,y_2,y_3)$, ${\bf z}=(z_1,z_2,z_3)$. Hence, $D=({\bf z}\cdot [{\bf x}\times {\bf y}])$. Similarly, the vectors ${\pmb \alpha }=(\alpha _1,\alpha _2,\alpha _3)$,
${\pmb \beta } =(\beta _1,\beta _2,\beta _3)$, ${\pmb \tau }=(\tau _1,\tau _2,\tau _3)$, ${\pmb \chi }=(\chi _1 ,\chi _2,\chi _3)$, ${\pmb \eta }=(\eta _1 ,\eta _2,\eta _3)$, ${\pmb \upsilon }=(\upsilon _1,\upsilon _2,\upsilon _3)$ are introduced. From \eqref{GrindEQ__21_}, \eqref{GrindEQ__26_} one has
$${\bf z}=X{\pmb \alpha }+Y{\pmb \beta },\  {\bf x}=-Y{\pmb \chi }+\dot{X}{\pmb \tau } -\left(1+\frac{Y^{2} }{R(1-e^{2} )} \right){\pmb \eta } +\left(X-\frac{3}{2} t\dot{X}\right){\pmb \upsilon },$$
$${\bf y}=X{\pmb \chi }+\dot{Y}{\pmb \tau +}\frac{XY}{R(1-e^{2} )}{\pmb \eta }+\left(Y-\frac{3}{2} t\dot{Y}\right){\pmb \upsilon }$$
This gives
$$D=(X\pmb \alpha +Y\pmb \beta )\left(\left(X\dot{X}+Y\dot{Y}\right)[\pmb \tau \times \pmb \chi ]+X[\pmb \chi \times \pmb \eta ]+\left(\frac{XY\dot{X}+Y^{2} \dot{Y}}{R(1-e^{2} )} +\dot{Y}\right)[\pmb \tau \times \pmb \eta ]+\right.$$
$$ +\left(\frac{3}{2} t(X\dot{X}+Y\dot{Y})-X^{2} -Y^{2} \right)[\pmb \chi \times \pmb \upsilon ]-(X\dot{Y}-Y\dot{X})[\pmb \tau \times \pmb \upsilon ]$$
$$\left. -\left(Y+\frac{Y^{3} }{R(1-e^{2} )} -\frac{3}{2} t\dot{Y}-\frac{3}{2} \frac{tY^{2} \dot{Y}}{R(1-e^{2} )} +\frac{X^{2} Y}{R(1-e^{2} )} -\frac{3}{2} \frac{tYX\dot{X}}{R(1-e^{2} )} \right)[\pmb \eta \times \pmb \upsilon ]\right)$$

In Appendix D it is shown that
\begin{equation} \label{GrindEQ__28_}
\frac{XY\dot{X}+Y^{2} \dot{Y}}{R(1-e^{2} )} +\dot{Y}=\frac{2e+X}{\sqrt{1-e^{2} } }
\end{equation}

Using \eqref{GrindEQ__6_}, \eqref{GrindEQ__9_}, \eqref{GrindEQ__12_}, (15) and \eqref{GrindEQ__28_}, we obtain
$$D(t)\equiv D(X,Y)=\left(\pmb \alpha X+\pmb \beta Y\right)\left([\pmb \chi \times \pmb \eta ]X+[\pmb \tau \times \pmb \chi ]\frac{eY}{\sqrt{1-e^{2} } } +[\pmb \tau \times \pmb \eta ]\frac{2e+X}{\sqrt{1-e^{2} } }+ \right. $$
$$+\left(\frac{3}{2} \frac{etY}{\sqrt{1-e^{2} } } -X^{2} -Y^{2} \right)[\pmb \chi \times \pmb \upsilon ]-\sqrt{1-e^{2} } [\pmb \tau \times \pmb \upsilon ]$$
\begin{equation} \label{GrindEQ__29_}
\left. -\left(2-\frac{eX}{(1-e^{2} )} \right)Y[\pmb \eta \times \pmb \upsilon ]+\frac{3}{2} \frac{t(2e+X)}{\sqrt{1-e^{2} } } [\pmb \eta \times \pmb \upsilon ]\right)
\end{equation}

That is, the volume can be represented as the modulus of a third-degree polynomial in the Cartesian coordinates of the reference orbit. The coefficients of this polynomial are functions of the initial relative coordinates and velocities of the deputies. They also explicitly linearly depend on time.

\noindent

\begin{center} \textit{2.6. Equal Orbital Periods of the Spacecraft}\end{center}

\noindent

The condition that all spacecraft have the same orbital period is equivalent to the condition of equality of energies $({\bf V}+{\bf u}_{m} )^{2} /2-1/\sqrt{({\bf R}+{\bf r}_{m} )^{2} } =V^{2} /2-1/R$. In the linear approximation, this yields
\begin{equation} \label{GrindEQ__30_}
({\bf Vu}_{m} )\approx -({\bf Rr}_{m} )/R^{3}
\end{equation}
It follows that for equal periods, the relative velocities in the plane of the reference orbit can be represented as
\begin{equation} \label{GrindEQ__31_}
\left[\begin{array}{c} {u_{xm} } \\ {u_{ym} } \end{array}\right]=-\frac{(Xx_{m} +Yy_{m} )}{R^{3} V^{2} } \left[\begin{array}{c} {\dot{X}} \\ {\dot{Y}} \end{array}\right]+w_{m} (t)\left[\begin{array}{c} {-\dot{Y}} \\ {\dot{X}} \end{array}\right]
\end{equation}
and can be described by a single function
\[w_m=(u_{ym} \dot{X}-u_{xm} \dot{Y})/V^{2} \]

It can be shown that \eqref{GrindEQ__30_} and \eqref{GrindEQ__31_} are equivalent to $\upsilon _m=0 $ in \eqref{GrindEQ__26_} (the semi-major axes of the orbits of all spacecraft are equal).  This imposes corresponding restrictions on the relative velocities $u_x, u_{y} $ at $t=t_{0} $. Substituting $u_{xm} ,u_{ym}$ from \eqref{GrindEQ__26_} at $\upsilon _m =0$ into the expression for $w_m$ the formula
$$w_m =\chi _m +\frac{X\dot{Y}-Y\dot{X}}{R^{3} V^{2} } \tau _m +\left(\left(\frac{X}{R(1-e^{2} )} -\frac{(X+e+X/R)}{R} \right)\dot{Y}\dot{X}+\frac{Y}{R^{2} } \left(\dot{X}\right)^{2} \right)\frac{\eta _{m} }{V^{2} } $$
is obtained. Using \eqref{GrindEQ__6_}, \eqref{GrindEQ__8_}, \eqref{GrindEQ__9_}, and (15) it is transformed to
$$w_m =\chi _m +\frac{\sqrt{1-e^{2} } }{R^{3} V^{2} } \tau _m +\left(1+\frac{(1-R)R}{(1-e^{2} )} \right)\frac{Y}{R^{3} V^{2} } \eta _m$$
Therefore, the following formula can be used instead of \eqref{GrindEQ__26_} in the case of equal periods
\begin{equation} \label{GrindEQ__32_}
\left[\begin{array}{c} {x_m } \\ {y_m } \\ {w_m } \end{array}\right]=\left[\begin{array}{ccc} {-Y} & {\dot{X}} & {-\left(1+\frac{Y^{2} }{R(1-e^{2} )} \right)} \\ {X} & {\dot{Y}} & {\frac{XY}{R(1-e^{2} )} } \\ {1} & {\frac{\sqrt{1-e^{2} } }{R^{3} V^{2} } } & {\left(1+\frac{(1-R)R}{1-e^{2} } \right)\frac{Y}{R^{3} V^{2} } } \end{array}\right]\left[\begin{array}{c} {\chi _m } \\ {\tau _m } \\ {\eta _m } \end{array}\right]
\end{equation}

\noindent {}

\begin{center} \textit{2.7. The Evolution of the Tetrahedron Volume in the Case of Identical Periods ($\pmb \upsilon =0$). Existence of a Formations with a Non-Vanishing Volume}\end{center}

\noindent {}

In an important special case of identical periods of all spacecraft $\pmb \upsilon =0$, the only case we will be focused on further, \eqref{GrindEQ__29_} takes the form
$$D(t)\equiv D (X,Y)=\left(\pmb \alpha X+\pmb \beta Y\right)\left([\pmb \chi \times \pmb \eta ]X+[\pmb \tau \times \pmb \chi ]\frac{eY}{\sqrt{1-e^{2} } } +[\pmb \tau \times \pmb \eta ]\frac{2e+X}{\sqrt{1-e^{2} } } \right)$$
The volume can be represented as a modulus of a second-degree polynomial in the Cartesian coordinates of the reference orbit with coefficients that depend on the initial relative coordinates and velocities
\begin{equation} \label{GrindEQ__33_}
{\rm V}(X,Y)=|c_1 X^2 +c_2 Y^2 +c_3 XY+c_4 X+c_5 Y|
\end{equation}
$$c_1 =\frac{(\pmb \alpha [\pmb \chi \times \pmb \eta ])}{6} +\frac{(\pmb \alpha [\pmb \tau \times \pmb \eta ])}{6\sqrt{1-e^{2} } } ,\ \ c_2 =\frac{e(\pmb \beta [\pmb \tau \times \pmb \chi ])}{6\sqrt{1-e^{2} } } $$
$$c_3 =\frac{(\pmb \beta [\pmb \chi \times \pmb \eta ])}{6} +\frac{(\pmb \beta [\pmb \tau \times \pmb \eta ])}{6\sqrt{1-e^{2} } } +\frac{e(\pmb \alpha [\pmb \tau \times \pmb \chi ])}{6\sqrt{1-e^{2} } } $$
$$c_4 =\frac{e(\pmb \alpha [\pmb \tau \times \pmb \eta ])}{3\sqrt{1-e^{2} } } , \ \ c_5 =\frac{e(\pmb \beta [\pmb \tau \times \pmb \eta ])}{3\sqrt{1-e^{2} } } $$

As shown in Appendix E, for any numbers $c_n$, $n=1,...,5$ it is possible to find non-zero vectors $\pmb \alpha ,\pmb \beta ,\pmb \chi ,\pmb \eta ,\pmb \tau $ yielding these coefficients. Moreover, the vectors $\pmb \alpha ,\pmb \beta ,\pmb \chi ,\pmb \eta ,\pmb \tau $ are not determined uniquely. Formations are possible in which all spacecraft are permanently in the same plane (all $c_n =0$), but not all of their orbits lie in the plane of the reference orbit ($\pmb \alpha $ and $\pmb \beta $ are different from zero and not parallel to each other).

Different formations can have the same dependence of the volume on time. For example, it is obvious that rotation $\pmb \alpha ,\pmb \beta ,\pmb \chi ,\pmb \eta ,\pmb \tau $ in the space $m = 1,2,3$ does not change the coefficients $c_{1-5} $. As shown in Appendix E, other transformations of $\pmb \alpha ,\pmb \beta ,\pmb \chi ,\pmb \eta ,\pmb \tau $ that do not change the behavior of the volume exist.

The lines of constant volume for arbitrary $X$ and $Y$ represent a family of curves of the 2nd order. The volume vanishes at the points where the curve V$(x,y) = 0 $ intersects the ellipse of the reference orbit. From \eqref{GrindEQ__33_} follows that the curve V$(x,y) = 0$  always passes through the origin.

If $\pmb \alpha $ is parallel to $\pmb \beta $ or/and $[\pmb \tau \times \pmb \chi ]$,  $[\pmb \chi \times \pmb \eta ]$, $[\pmb \tau \times \pmb \eta ]$ are parallel to each other, then the curves V=const are hyperbolas, and the lines ${\rm V}(x,y)=0$ are 2 straight lines. One line passes through the center of coordinates, providing at least 2 intersections with the ellipse of the reference orbit. Another line may intersect or do not intersect the reference orbit, may touch it, or coincide with the first straight line. Thus, the volume can vanish 2, 3, or 4 times. According to \eqref{GrindEQ__27_}, the condition of parallelism $\pmb \alpha $ and $\pmb \beta $ is equivalent to the condition $u_{zm} =Cz_{m} $ at the initial moment of time. The value $C$ does not depend on $m$. If this condition is satisfied at one of the moments of time, then it will be satisfied at all moments of time, but with a different  $C$.

If the conditions of parallelism $\pmb \alpha $ and $\pmb \beta $ and mutual parallelism $[\pmb \tau \times \pmb \chi ]$, $[\pmb \chi \times \pmb \eta ]$ and  $[\pmb \tau \times \pmb \eta ]$ are not satisfied, then the curve V = 0 can be, in particular, an ellipse passing through the center of coordinates. If the size of this ellipse is small enough, then it will be entirely inside the reference orbit.  Then the volume of the tetrahedron will not be zeroed. Since 2 ellipses can intersect at no more than 4 points (except the case they are coincide), the volume can be zeroed no more than 4 times per period.

 The curve ${\rm V}=0$ can also correspond to a parabola passing through the center of coordinates, which provides the number of intersections with the ellipse of the reference orbit from 2 to 4 times.

\noindent {}

\begin{center} \textit{2.8. Start at Perihelion}\end{center}

\noindent {}

The relationship between vectors $\pmb \alpha $, $\pmb \beta $, $\pmb \chi $, $\pmb \eta $, $\pmb \tau $ and relative coordinates, and velocities at perihelion (and similarly at aphelion) is relatively simple. In this case $X=R_p$, $\dot{Y}=V_p $, $Y=0$, $\dot{X}=0$. The formula $R_p^{-3} V_p^{-2} \sqrt{1-e^{2} } =R_p^{-2} V_p^{-1} $ can be derived from \eqref{GrindEQ__3_}. Taking this into account the following expression can be obtained from \eqref{GrindEQ__32_}
$$\left[\begin{array}{c} {x_{0m} } \\ {y_{0m} } \\ {w_{0m} } \end{array}\right]=\left[\begin{array}{ccc} {0} & {0} & {-1} \\ {R_{p} } & {V_{p} } & {0} \\ {1} & {(R_{p}^{2} V_{p} )^{-1} } & {0} \end{array}\right]\left[\begin{array}{c} {\chi _m } \\ {\tau _m } \\ {\eta _m } \end{array}\right]$$
Here and below, the index 0 refers to quantities at perihelion. Taking $R_p $ and $V_p$ expressed in terms of the eccentricity into account, this gives
\begin{equation} \label{GrindEQ__34_}
\pmb \eta =-{\bf x}_{0} , \pmb \chi =\frac{1+e}{e} {\bf w}_{0} -\frac{{\bf y}_{0} }{eR_p } , \pmb \tau =\frac{1+e}{e} \cdot \frac{{\bf y}_0 -R_p {\bf w}_{0} }{V_p } =\frac{R_p V_p }{e} ({\bf y}_0 -R_{p} {\bf w}_0 )
\end{equation}
In addition, \eqref{GrindEQ__27_} gives
$$\pmb \alpha ={\bf z}_{0} /R_p,\ \ \pmb \beta ={\bf u}_{z0}/V_p$$
Accordingly, the volume \eqref{GrindEQ__33_} is a linear combination of mixed products
$({\bf z}_0 \left[{\bf x}_0 \times {\bf y}_0 \right])$, $({\bf z}_0 \left[{\bf x}_0 \times {\bf w}_0 \right])$,  $({\bf z}_0 \left[{\bf y}_0 \times {\bf w}_0 \right])$,  $({\bf u}_{z0} \left[{\bf x}_0 \times {\bf y}_0 \right])$, $({\bf u}_{z0} \left[{\bf x}_0 \times {\bf w}_0 \right])$, $({\bf u}_{z0} \left[{\bf y}_0 \times {\bf w}_0 \right])$. The coefficients $c_{1-5} $ take the form

$$c_{1} =\frac{1}{3R_p } ({\bf z}_{0} [{\bf x}_0 \times {\bf w}_0 ])-\frac{({\bf z}_0 [{\bf x}_0 \times {\bf y}_0 ])}{6R_p^{2} },\ \ c_{2} =\frac{({\bf u}_{z0} [{\bf y}_0 \times {\bf w}_0 ])}{6V_p }$$
\begin{equation} \label{GrindEQ__35_}
c_{3} =\frac{({\bf u}_{z0} [{\bf x}_0 \times {\bf w}_0 ])}{3V_{p} } -\frac{({\bf u}_{z0} [{\bf x}_0 \times {\bf y}_0 ])}{6V_p R_p } +\frac{({\bf z}_0 [{\bf y}_0 \times {\bf w}_0 ])}{6R_p }
\end{equation}
$$c_{4} =\frac{({\bf z}_0 [{\bf x}_0 \times {\bf y}_0 ])}{3R_p } -\frac{({\bf z}_0 [{\bf x}_0 \times {\bf w}_0 )])}{3} ,   c_{5} =\frac{({\bf u}_{z0} [{\bf x}_0 \times {\bf y}_0 ])}{3V_p } -\frac{R_p }{3V_p } ({\bf u}_{z0} [{\bf x}_0 \times {\bf w}_0 )])$$

The substitution of \eqref{GrindEQ__34_} into \eqref{GrindEQ__32_} gives
$$x_{m} =\left(\frac{Y}{eR_p } +\frac{R_p V_p }{e} \dot{X}\right)y_{0m} +\left(1+\frac{Y^{2} }{R(1-e^{2} )} \right)x_{0m} -\left(\frac{1+e}{e} Y+\frac{R_p^2 V_p }{e} \dot{X}\right)w_{0m}$$
 $$y_m =-\frac{X}{eR_p } y_{0m} +\frac{R_p V_p \dot{Y}}{e} y_{0m} -\frac{XY}{R(1-e^{2} )} x_{0m} +\frac{X(1+e)-R_p^{2} V_{p} \dot{Y}}{e} w_{0m} $$

After performing simple yet cumbersome calculations, using equations \eqref{GrindEQ__3_}, \eqref{GrindEQ__6_}, (8--13), and (15), the following formulas are obtained
\begin{equation} \label{GrindEQ__36_}
x_m =S_{xx} x_{0m} +S_{xy} y_{0m} +S_{xw} w_{0m}
\end{equation}
$$S_{xx} =1+\frac{Y^{2} }{R(1-e^{2} )} ,\ \  S_{xy} =\frac{(R_{p} -X)Y}{R_{p} R} , \ \S_{xw} =-\frac{(R+R_{p} -X)Y}{R} $$
\begin{equation} \label{GrindEQ__37_}
y_m =S_{yx} x_{0m} +S_{yy} y_{0m} +S_{yw} w_{0m}
\end{equation}
$$S_{yx} =-\frac{XY}{R(1-e^{2} )},\ \ S_{yy} =\frac{X^{2} +(1-e^{2} )(R_{p} -X)}{R R_p },\ \ S_{yw} =-\frac{(1+e)(R_p -X)^{2} }{R} $$
$$z_{m} =(X/R_p )z_{0m} +(Y/V_p )u_{z0m}$$

Note that there are no singularities in expressions (35--37) for zeroed $e$.

The dependence of the coefficients in expressions \eqref{GrindEQ__36_}, \eqref{GrindEQ__37_} on the position in the reference orbit (on the coordinate $X$ of the chief) has the following properties (Fig. 2).

\begin{figure}
\centering {\includegraphics*[width=12cm]{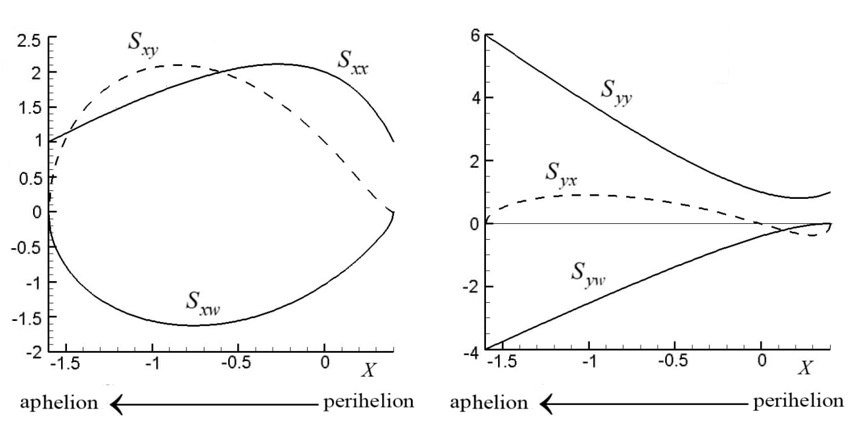}}
\caption*{\textbf{Fig. 2.} Dependencies of the coefficients in formulas \eqref{GrindEQ__36_}, \eqref{GrindEQ__37_} on $X$ for $Y\ge 0$ (movement from perihelion to aphelion). $e=0.6$.}
\end{figure}

$S_{xx} $ increases from 1 at perihelion to a maximum, then drops again to 1 at aphelion. For $e=0.6$ $\max S_{xx}\approx 2.11$ is reached at the point $X\approx -0.265$. At the point $X=-e$ ($Y=\sqrt{1-e^2} =\max $, $R=1$) $S_{xx} $=2 for any eccentricities.

$S_{xy} $ has the same sign as $Y$. It increases from 0 at the perihelion to a maximum $\approx 2.105$ at the point $X\approx -0.85$ (at $e=0.6$). Then $S_{xy} $ falls back to 0 at the aphelion. At the point $X=-e$ $S_{xy} =\sqrt{(1+e)/(1-e)} =V_p$. At this point at $e=0.6$ $S_{xy} =2$, what coincides with $S_{xx}$ at this point.

$S_{xw}$ is negative when moving from perihelion to aphelion ($Y\ge 0$) and positive otherwise. At aphelion and perihelion $S_{xw} =0$. At $e=0.6$ the minimum $S_{xw} $ is reached at the point $X\approx -0.76$ and is approximately equal to $-1.625$ (at $Y\ge 0$).

$S_{yy} >0$ does not depend on the sign of $Y$.  This coefficient is equal to 1 at the perihelion $X=X_p$ and at $X=0$. $S_{yy} $ reaches a minimum in the interval $0<X<X_p$.  At $e=0.6$ the minimum is reached at the point $X\approx -0.225$ and is equal to $\approx 0.805$, i.e. differs little from 1. At $X<-0.225$ on the way to the aphelion $S_{yy} $ increases, reaching a value $\left(4-(1-e)^{2} \right)(1-e^{2} )^{-1} \ge 3$ at the aphelion, which is approximately equal to 6 at $e=0.6$. At $X=-e$ $S_{yy} =(1-e)^{-1} $.

$S_{yw} $ on the way from perihelion to aphelion monotonically decreases from 0 to $-4$.

$S_{yx} $ is proportional to $-XY$. It has different signs in different quarters of the reference orbit. On the way from perihelion to aphelion $S_{yx}$ firstly decreases, reaching a minimum $\approx -0.355$ at the point $X\approx 0.3$, then increases to a maximum $\approx 0.924$ at the point $X\approx -1.015$. Here all values are given for $e=0.6$.

As Fig. 2 shows, at $e = 0.6$ the greatest change during motion along the orbit is undergone by the value $S_{yy} $, and the least by $S_{yx} $. The dependencies $S_{xy} $ and $-S_{xw} $ on $X$ are qualitatively similar to each other, as well as $S_{yx}$ and $1-S_{yy} $.

Note, that the condition of equality of periods \eqref{GrindEQ__30_} gives a specific value of the $y$-component of the relative velocity $u_{y0m} =-x_{0m} /(R_p^{2} V_p)$ at perihelion; the $x$-component of the relative velocity can be arbitrary.

\noindent {}

\begin{center}
\textbf{3. Results of Calculations of the Evolution of the Tetrahedron}
\end{center}

\begin{center}
\textit{3.1. Examples of the Evolution of the Volume and Quality of the Tetrahedrons}
\end{center}

\noindent

Let's consider several examples of the evolution of the volume and another important value, characterizing how the tetrahedron is close to a regular one, -- the quality $Q$ of the tetrahedron. The quality of the tetrahedron affects the accuracy of the measurements [26-28]. It can be determined in different ways [28-32]. In the presented paper the simplest formula $Q=12(3{\rm V})^{2/3} /L^{2} $ [2, 3, 33, 34] will be used. Here $L^{2} $ is the sum of the squared lengths of all tetrahedron's edges. According to this formula, for a regular tetrahedron $Q=1$. Any deviation from a regular tetrahedron leads to decrease in $Q$.

By virtue of the linearity of the used approximation, multiplying all relative coordinates and velocities by the same number results in the vectors ${\bf w}$, $\pmb \alpha $, $\pmb \beta$, $\pmb \chi $, $\pmb \tau$, $\pmb \eta$ to be multiplied by this number and the volume and coefficients $c_{1-5}$ to be multiplied by the cube of this number. Therefore, the specific size of the tetrahedron is not important, including if it is of the order of 1. If any relative distance is equal to 1 in terms of the dimensionless variables used in the present work, and the dimensionless relative velocity is $u$, and the distance should be equal to $r_0$ meters in dimensional quantities, then we need  necessary to multiply all distances by $r_0$ and all velocities by $2\pi r_0 /T$ in order to obtain dimensional relative quantities. For  $r_0=10^{6}$ m and $a=1$ AU  $2\pi r_0 /T=0.2$ m/s.

On the left panels in Figures 3-9 the time dependence of the ratio ${\rm V}/{\rm V}_*$ is depicted, where ${\rm V}_{*} $ is the volume of the tetrahedron at the moment of launch; on the central panel the time dependence of the quality of the tetrahedron \textit{Q} is depicted; on the right panel the ellipse of the reference orbit (solid bold line), isolines ${\rm V}(X,Y)=0$ (dash bold lines), and isolines ${\rm V}/{\rm V}_{*} {\rm =}1$ (dash-dot lines) are depicted. The regions ${\rm V}\le {\rm V}_{*} $ are shaded. All figures correspond to the case of identical periods $\upsilon _m =0$. Respectively, the relative velocities in the orbital plane $u_{xm} ,u_{ym} $ satisfy formula \eqref{GrindEQ__31_} and are determined by a single value $w_m$. Dimensionless quantities are used.  The options are as follows.

Fig. 3. Eccentricity $e=0.6$. Start at perihelion ($t_0=0$). At the start, the z-components of the relative velocities $u_{z0} =0$ and the value $w_0 =0$. The relative coordinates correspond to a regular tetrahedron with a unit edge and are equal to
$$\left[\begin{array}{c} {{\bf x}_{0} } \\ {{\bf y}_{0} } \\ {{\bf z}_{0} } \end{array}\right]=\left[\begin{array}{ccc} {1/\sqrt{3} } & {-1/\left(2\sqrt{3} \right)} & {-1/\left(2\sqrt{3} \right)} \\ {0} & {-1/2} & {1/2} \\ {\sqrt{2/3} } & {\sqrt{2/3} } & {\sqrt{2/3} } \end{array}\right]$$
The relative velocities at the start and the integration constants are

$$\left[\begin{array}{c} {{\bf u}_{x0} } \\ {{\bf u}_{y0} } \\ {{\bf u}_{z0} } \end{array}\right]=\left[\begin{array}{ccc} {0} & {0} & {0} \\ {-1.804} & {0.9021} & {0.9021} \\ {0} & {0} & {0} \end{array}\right]$$

$$ \left[\begin{array}{l} {\pmb \alpha } \\ \pmb {\beta } \\ \pmb {\chi } \\ \pmb {\tau } \\ \pmb {\eta } \end{array}\right]=\left[\begin{array}{ccc} {\begin{array}{c} {2.041} \\ {0} \end{array}} & {\begin{array}{l} {2.041} \\ {{\rm \; \; \; \; }0} \end{array}} & {\begin{array}{l} {2.041} \\ {{\rm \; \; \; \; }0} \end{array}} \\ {0} & {2.083} & {-2.083} \\ {\begin{array}{c} {0} \\ {-0.5774} \end{array}} & {\begin{array}{c} {-2/3} \\ {0.2887} \end{array}} & {\begin{array}{c} {2/3} \\ {0.2887} \end{array}} \end{array}\right]$$
The expression for the volume at $u_{z0} =0$ and $w_0=0$ has the form  $${\rm V}(X,Y)=({\bf z}_0 [{\bf x}_0 \times {\bf y}_0])\left|(X/R_p)^{2} -2(X/R_p)\right|/6$$
 Accordingly,   ${\rm V/V}_* =\left|(X/R_p)^{2} -2(X/R_p)\right|$. The volume reaches its maximum at aphelion and turns to 0 twice per period. The dependence of the volume on time is the same in half-periods. Since the mixed product $({\bf z}_0 [{\bf x}_0 \times {\bf y}_0])$ in space $m=1,2,3$ is invariant with respect to the rotation of vectors in ordinary space, the behavior of the volume does not depend on the orientation of the tetrahedron at perihelion. Indeed, rotation around the chief, for example, around the $z$ axis by an angle of $\varphi $ means replacing ${\bf x}_0 \to {\bf x}_{0} \cos \varphi +{\bf y}_0 \sin \varphi $, ${\bf y}_0 \to {\bf y}_0 \cos \varphi -{\bf x}_0 \sin \varphi $, ${\bf z}_0 \to {\bf z}_0$. This transformation does not change $({\bf z}_{0} [{\bf x}_0 \times {\bf y}_0])$.  Furthermore, imposing restrictions of the form of parallelism ${\bf u}_{z0} $ and ${\bf z}_0 $, and ${\bf w}_0$ and ${\bf y}_0$, means that the volume will depend only on $({\bf z}_0 [{\bf x}_0 \times {\bf y}_0])$ and will be invariant with respect to the rotation of the tetrahedron in ordinary space at perihelion. The restrictions above are equivalent to $u_{zm} =C_z z_m$ and  $u_{x0m} =-V_p w_{0m} =C_x y_{0m} $, where  $C_z ,C_x$ are constants independent of $m$.

\begin{figure}
\centering {\includegraphics*[width=14cm]{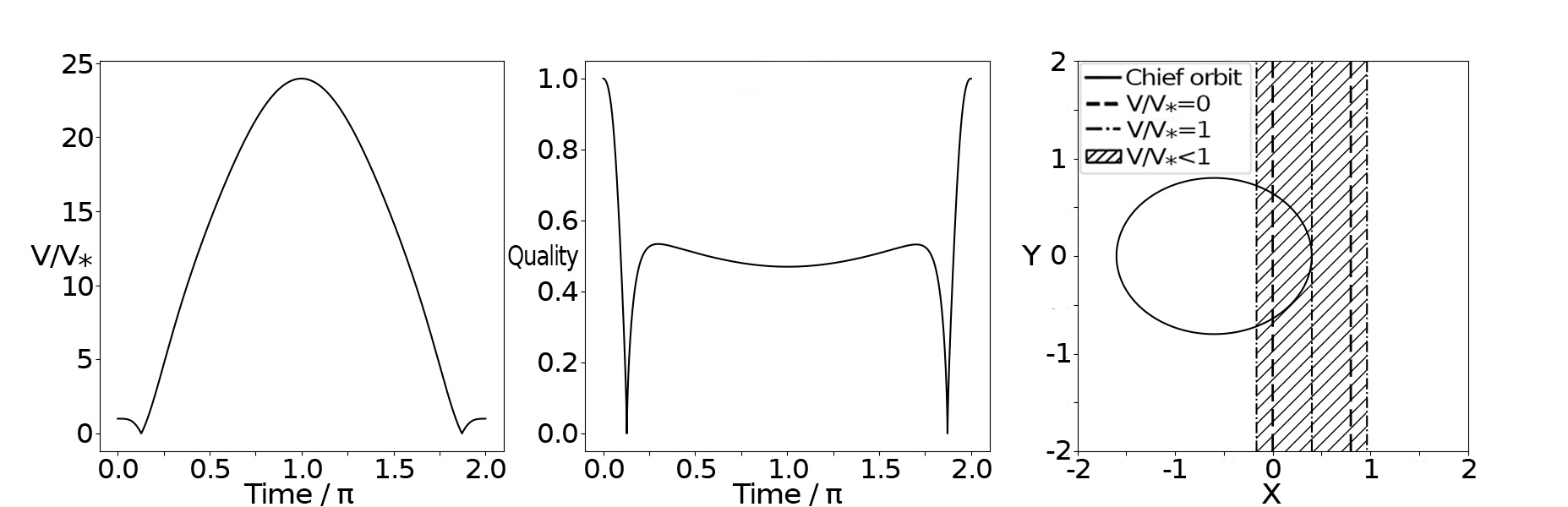}}
\caption*{\textbf{Fig. 3. }Dependence ${\rm V}(t)/{\rm V}_{*} $ (left), $Q(t)$(center), and ${\rm V}(X,Y)/{\rm V}_{*} $ (right) in the case of $e= 0.6$. At perihelion ($t=0$) the tetrahedron is regular and ${\bf u}_{z} =0$, ${\bf w}=0$.}
\end{figure}

\noindent

Calculations show that the time-dependence of quality $Q$ of the tetrahedron with ${\bf u}_{z} =0$, ${\bf w}=0$ is also invariant with respect to tetrahedron rotation around the chief at perihelion. At least this statement is true for a large number of variants of such rotations. It was not possible to explain this analytically yet. $Q$ turns to 0 at the same moments of time as the volume. Note that over a large period of time, corresponding to $X<0$,  tetrahedron quality changes little and it is quite high. We emphasize that the specific shape of the tetrahedron depends on its initial orientation, that is opposite to ${\rm V}(t)$ and $Q(t)$.

Fig. 4. The eccentricity and starting values ${\bf x}_0$, ${\bf y}_0$, ${\bf z}_{0}$, ${\bf u}_{z0}$, ${\bf w}_0$ are the same as in Fig. 3, but the start occurs at the aphelion $t_0 =\pi $. In this case
$$\left[\begin{array}{c} {{\bf u}_{x0} } \\ {{\bf u}_{y0} } \\ {{\bf u}_{z0} } \end{array}\right]=\left[\begin{array}{ccc} {0} & {0} & {0} \\ {-0.451} & {0.2255} & {0.2255} \\ {0} & {0} & {0} \end{array}\right], \left[\begin{array}{c} {\pmb \alpha } \\ {\pmb \beta } \\ {\pmb \chi } \\ {\pmb \tau } \\ {\pmb \eta } \end{array}\right]=\left[\begin{array}{ccc} {\begin{array}{c} {-0.51} \\ {0} \end{array}} & {\begin{array}{c} {-0.51} \\ {0} \end{array}} & {\begin{array}{c} {-0.51} \\ {0} \end{array}} \\ {0} & {0.521} & {-0.521} \\ {\begin{array}{c} {0} \\ {-0.577} \end{array}} & {\begin{array}{c} {-2/3} \\ {0.289} \end{array}} & {\begin{array}{c} {2/3} \\ {0.289} \end{array}} \end{array}\right],$$
$${\rm V}/{\rm V}_{*} =\left|X^{2} /R_{a}^{2} +2X/R_{a} \right|.$$

Here $R_{a} =1+e$ is the radius of the reference orbit at aphelion. The ratio of volumes at perihelion and aphelion is smaller in this case and the quality of the tetrahedron is lower than for the case with start at perihelion. Evolution of V and \textit{Q} is invariant with respect to the rotation of the tetrahedron at the start point.

\begin{figure}
\centering {\includegraphics*[width=14cm]{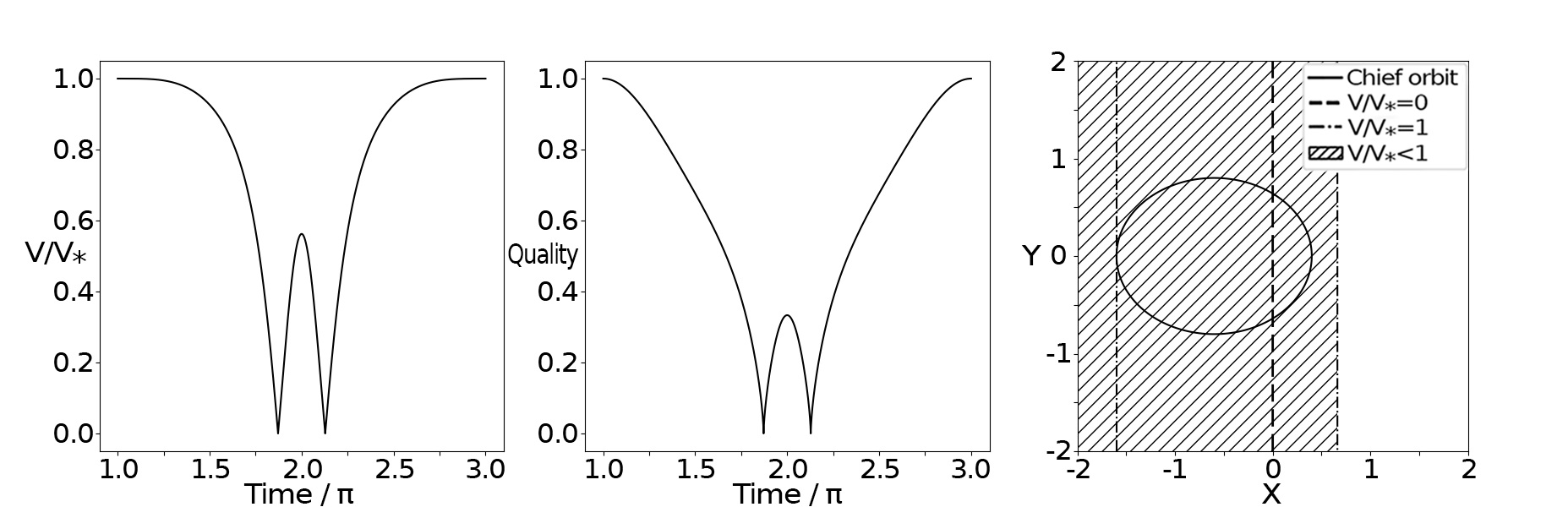}}
\caption*{\textbf{Fig. 4. } The same as Fig. 3, except for the start is at aphelion ($t_0 =\pi $).}
\end{figure}

\noindent{}

Fig. 5.  Eccentricity $e= 0.3$. Start at perihelion ($t_0=0$). The starting values ${\bf x}_0$, ${\bf y}_0$, ${\bf z}_0$, ${\bf u}_{z0}$, ${\bf w}_0$ are the same as in Fig. 3. In this case
$$\left[\begin{array}{c} {{\bf u}_{x0} } \\ {{\bf u}_{y0} } \\ {{\bf u}_{z0} } \end{array}\right]=\left[\begin{array}{ccc} {0} & {0} & {0} \\ {-0.865} & {0.432} & {0.432} \\ {0} & {0} & {0} \end{array}\right], \left[\begin{array}{c} {\pmb \alpha } \\ {\pmb \beta } \\ {\pmb \chi } \\ {\pmb \tau } \\ {\pmb \eta } \end{array}\right]=\left[\begin{array}{ccc} {\begin{array}{c} {1.166} \\ {0} \end{array}} & {\begin{array}{c} {1.166} \\ {0} \end{array}} & {\begin{array}{l} {1.166} \\ {0} \end{array}} \\ {0} & {2.38} & {-2.38} \\ {\begin{array}{c} {0} \\ {-0.577} \end{array}} & {\begin{array}{c} {-1.59} \\ {0.287} \end{array}} & {\begin{array}{c} {1.59} \\ {0.287} \end{array}} \end{array}\right],$$
$${\rm V}/{\rm V}_{*} =\left|X^{2} /R_{p}^{2} -2X/R_{p} \right|,\ R_{p} =0.7.$$

\begin{figure}
\centering {\includegraphics*[width=14cm]{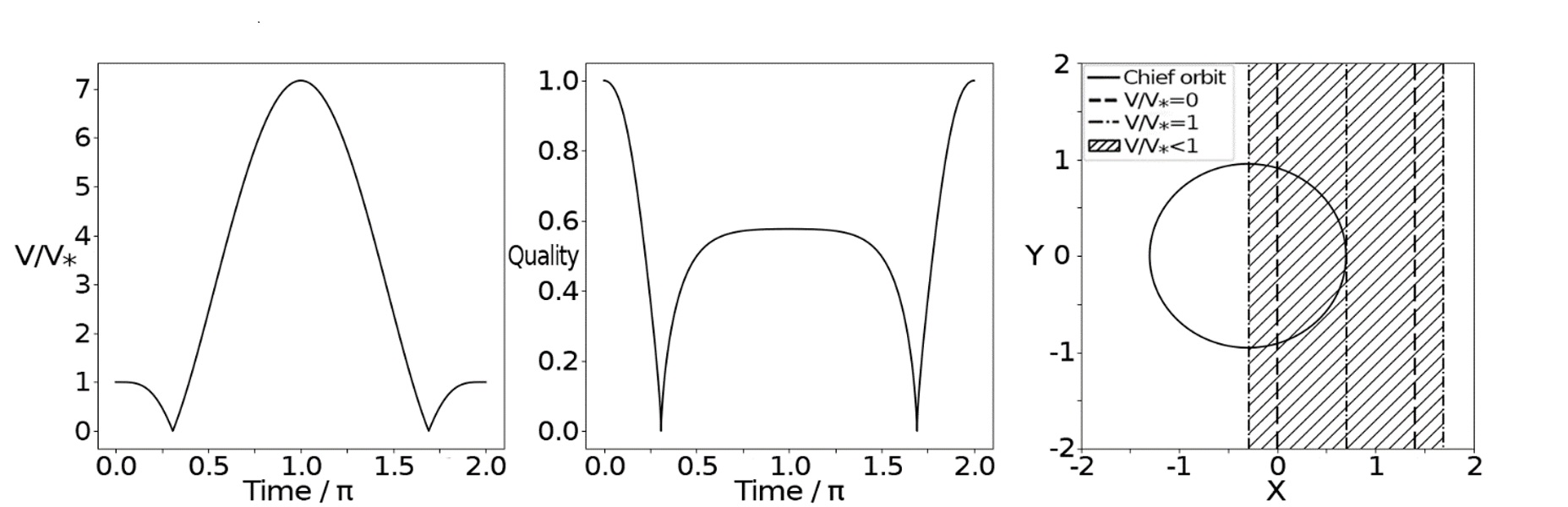}}
\caption*{\textbf{Fig. 5. } The same as Fig. 3, except for $e= 0.3$.}
\end{figure}

Fig. 6. Eccentricity $e = 0.9$. Start at perihelion ($t_0=0$). The starting values ${\bf x}_0$, ${\bf y}_0$, ${\bf z}_0$, ${\bf u}_{z0}$, ${\bf w}_0$ are the same as in Fig. 3. In this case
$$\left[\begin{array}{c} {{\bf u}_{x0} } \\ {{\bf u}_{y0} } \\ {{\bf u}_{z0} } \end{array}\right]=\left[\begin{array}{ccc} {0} & {0} & {0} \\ {-13.25} & {6.623} & {6.623} \\ {0} & {0} & {0} \end{array}\right], \left[\begin{array}{c} {\pmb \alpha } \\ {\pmb \beta } \\ {\pmb \chi } \\ {\pmb \tau } \\ {\pmb \eta } \end{array}\right]=\left[\begin{array}{ccc} {\begin{array}{c} {8.165} \\ {0} \end{array}} & {\begin{array}{c} {8.165} \\ {0} \end{array}} & {\begin{array}{l} {8.165} \\ {0} \end{array}} \\ {0} & {5.556} & {-5.556} \\ {\begin{array}{c} {0} \\ {-0.577} \end{array}} & {\begin{array}{c} {-0.242} \\ {0.289} \end{array}} & {\begin{array}{c} {0.242} \\ {0.289} \end{array}} \end{array}\right],$$
 $${\rm V}/{\rm V}_{*} =\left|X^{2} /R_{p}^{2} -2X/R_{p} \right|, R_{p} =0.1.$$

\begin{figure}
\centering {\includegraphics*[width=14cm]{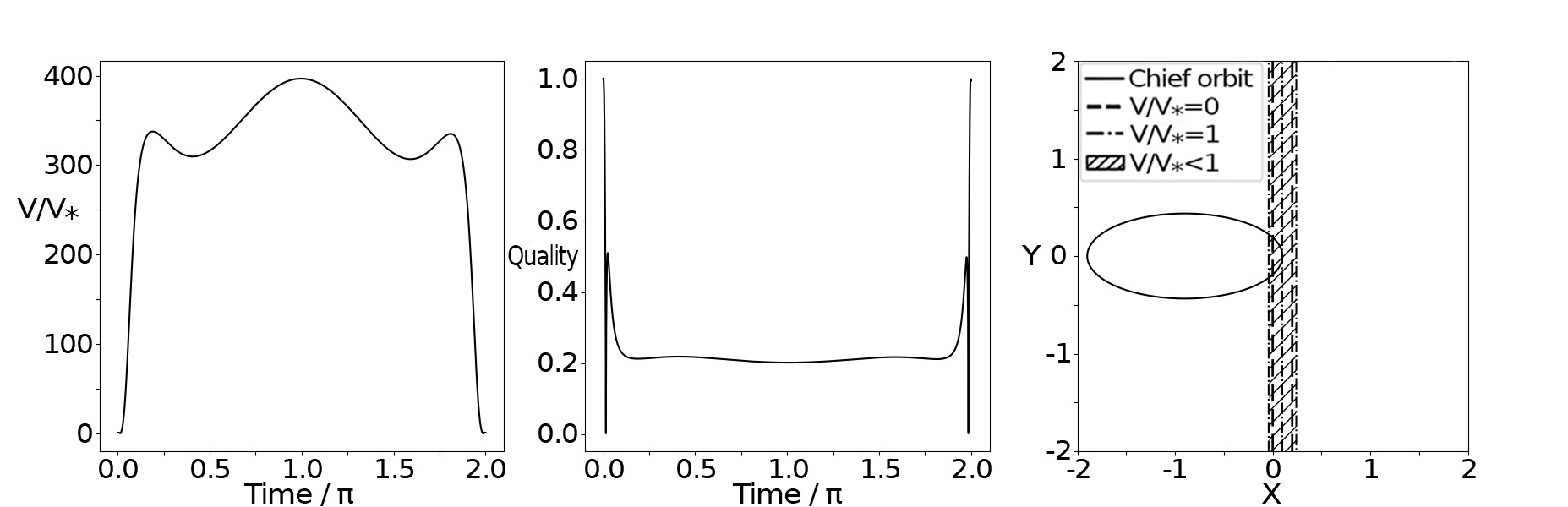}}
\caption*{\textbf{Fig.6. } The same as Fig. 3, except for $e= 0.9$.}
\end{figure}

\noindent

The comparison of Fig.-s 3, 5, 6 shows, that an increase in eccentricity leads to an increase in the ratio of volumes at aphelion and perihelion, as well as to an increase in the length of the sections of little-changing dependencies of volume and quality on time. The quality of the tetrahedron decreases in these sections as eccentricity increases.

Fig. 7.\textbf{ }An example of volume that vanishes four times per the period. In this variant, all initial parameters of the formation coincide with those in Fig. 3, except for the vector $w_{0} $, which is ${\bf w}_{0} =(14.96,0,0)$. This gives
$$\left[\begin{array}{c} {{\bf u}_{x0} } \\ {{\bf u}_{y0} } \\ {{\bf u}_{z0} } \end{array}\right]=\left[\begin{array}{ccc} {-29.92} & {0} & {0} \\ {-1.804} & {0.9021} & {0.9021} \\ {0} & {0} & {0} \end{array}\right],  \left[\begin{array}{c} {\pmb \alpha } \\ {\pmb \beta } \\ {\pmb \chi } \\ {\pmb \tau } \\ {\pmb \eta } \end{array}\right]=\left[\begin{array}{ccc} {\begin{array}{c} {2.041} \\ {0} \end{array}} & {\begin{array}{l} {2.041} \\ {{\rm \; \; \; \; }0} \end{array}} & {\begin{array}{l} {2.041} \\ {{\rm \; \; \; \; }0} \end{array}} \\ {39.89} & {2.083} & {-2.083} \\ {\begin{array}{c} {-7.979} \\ {-0.5774} \end{array}} & {\begin{array}{c} {-2/3} \\ {0.2887} \end{array}} & {\begin{array}{c} {2/3} \\ {0.2887} \end{array}} \end{array}\right],$$
 $${\rm V/V}_{*} \approx |6.25X^{2} +43.2XY-5X|.$$

In this case, the quality of the tetrahedron is low for the most part of the orbit. Additionally, there is no symmetry in the half-periods. Note, that in the dimensional variables, for a tetrahedron with the edge length of 1000 km, the value $w_1= 14.96$ corresponds to the velocity $u_{x1} =6$ m/s.

\begin{figure}
\centering {\includegraphics*[width=14cm]{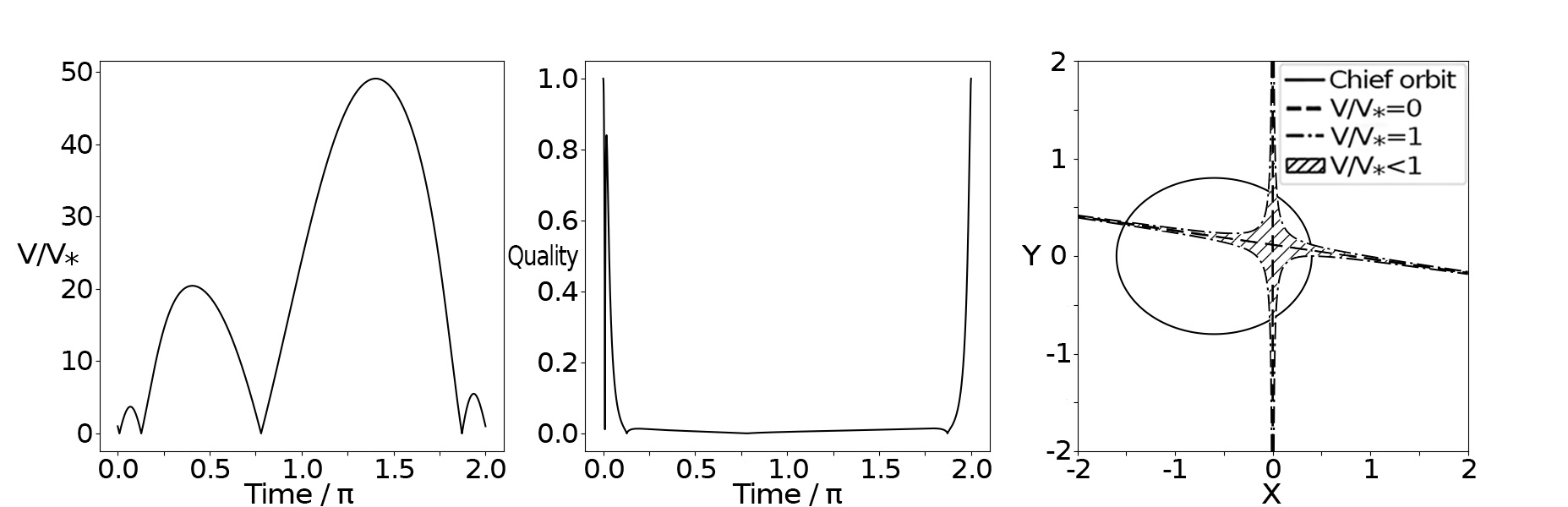}}
\caption*{\textbf{Fig.7. } An example of tetrahedron with volume zeroed 4 times.  Data are the same as on Fig. 3, except for ${\bf w}_0= [14.96, 0, 0]$.}
\end{figure}

Examples of tetrahedrons with non-vanishing volume are presented in Fig.-s 8, 9.

Fig. 8. Eccentricity $e= 0.6$. Start at perihelion ($t_0 = 0$). At the moment of start
$$\left[\begin{array}{c} {{\bf x}_0 } \\ {{\bf y}_0} \\ {{\bf z}_0 } \end{array}\right]=\left[\begin{array}{ccc} {0} & {1} & {0} \\ {40} & {0} & {2} \\ {16} & {0} & {0} \end{array}\right],{\bf w}_{0} =[100,\, 0,\, 3.125], {\bf u}_{z0} =[0,2,0].$$
This gives
$$\left[\begin{array}{c} {{\bf u}_{x0} } \\ {{\bf u}_{y0} } \\ {{\bf u}_{z0} } \end{array}\right]=\left[\begin{array}{ccc} {-200} & {0} & {-6.25} \\ {0} & {-3.125} & {0} \\ {0} & {2} & {0} \end{array}\right], \left[\begin{array}{c} {\pmb \alpha } \\ {\pmb \beta } \\ {\pmb \chi } \\ {\pmb \tau } \\ {\pmb \eta } \end{array}\right]=\left[\begin{array}{ccc} {\begin{array}{c} {40} \\ {0} \end{array}} & {\begin{array}{c} {0} \\ {{\rm 1}} \end{array}} & {\begin{array}{l} {0} \\ {0} \end{array}} \\ {100} & {0} & {0} \\ {\begin{array}{c} {0} \\ {0} \end{array}} & {\begin{array}{c} {0} \\ {-1} \end{array}} & {\begin{array}{c} {1} \\ {0} \end{array}} \end{array}\right],$$
$$ {\rm V/V}_{*} \approx |1.56X^{2} +2.344Y^{2} +1.875X|$$

Fig. 9. Eccentricity $e = 0.6$. Start at perihelion ($t_0=0$). At the moment of start
$$\left[\begin{array}{c} {{\bf x}_{0} } \\ {{\bf y}_{0} } \\ {{\bf z}_{0} } \end{array}\right]=\left[\begin{array}{ccc} {0} & {1} & {0} \\ {40} & {0} & {2} \\ {0.4} & {0} & {0} \end{array}\right],\ {\bf w}_0 =[100,\, 0,\, 3.125], {\bf u}_{z0} =[0,\, 2,\, 0].$$
That is, Fig. 9 differs from Fig. 8 only in the $z$-coordinate of the first spacecraft. This gives
$$\left[\begin{array}{c} {{\bf u}_{x0} } \\ {{\bf u}_{y0} } \\ {{\bf u}_{z0} } \end{array}\right]=\left[\begin{array}{ccc} {-200} & {0} & {-6.25} \\ {0} & {-3.125} & {0} \\ {0} & {2} & {0} \end{array}\right], \left[\begin{array}{c} {\pmb \alpha } \\ {\pmb \beta } \\ {\pmb \chi } \\ {\pmb \tau } \\ {\pmb \eta } \end{array}\right]=\left[\begin{array}{ccc} {\begin{array}{c} {1} \\ {0} \end{array}} & {\begin{array}{c} {0} \\ {{\rm 1}} \end{array}} & {\begin{array}{l} {0} \\ {0} \end{array}} \\ {100} & {0} & {0} \\ {\begin{array}{c} {0} \\ {0} \end{array}} & {\begin{array}{c} {0} \\ {-1} \end{array}} & {\begin{array}{c} {1} \\ {0} \end{array}} \end{array}\right],$$
$${\rm V/V}_* \approx |1.56X^{2} +93.75Y^{2} +1.875X|$$

In the case of Fig. 8 the ellipse of the curve V/V$_*=1$ practically coincidence with the ellipse of the reference orbit. This explains the small change in volume over time. In the case of Fig. 9, the ellipse V/V$_*=1$ is significantly flattened. As a result, the change in volume is large. In both cases, the quality of the tetrahedron is low at all points of the orbit.

\begin{figure}
\centering {\includegraphics*[width=14cm]{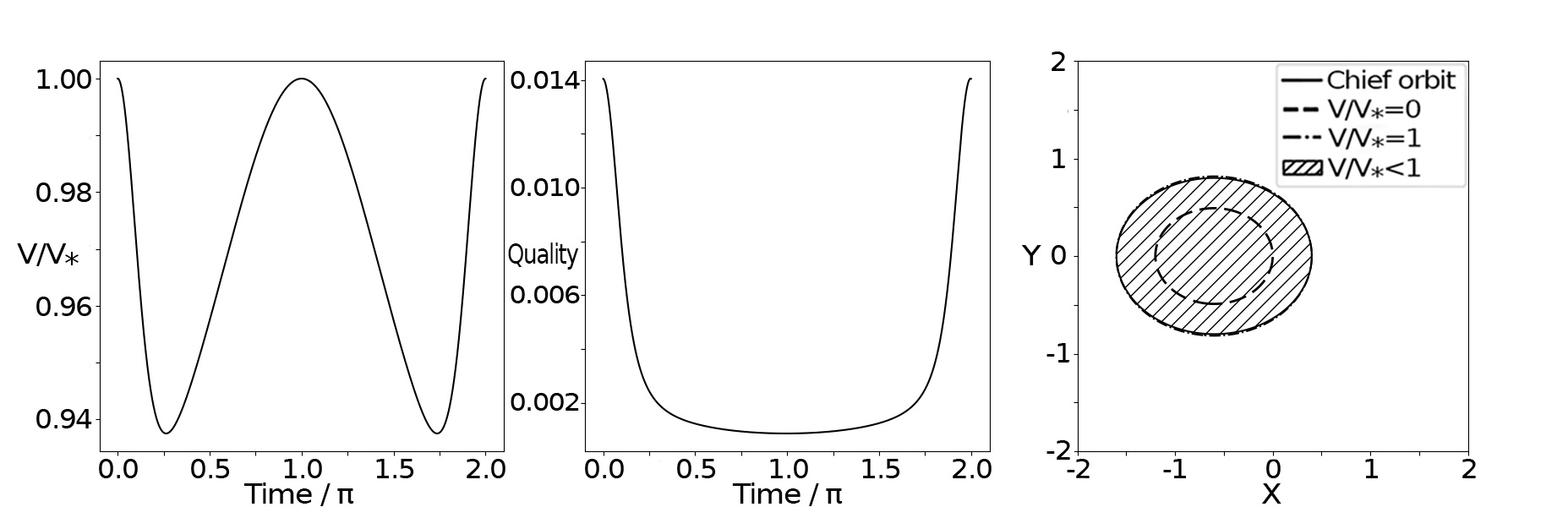}}
\caption*{\textbf{Fig.8. } An example of small changing, not zeroed volume. Data see in the text.}
\end{figure}

\noindent

\begin{figure}
\centering {\includegraphics*[width=14cm]{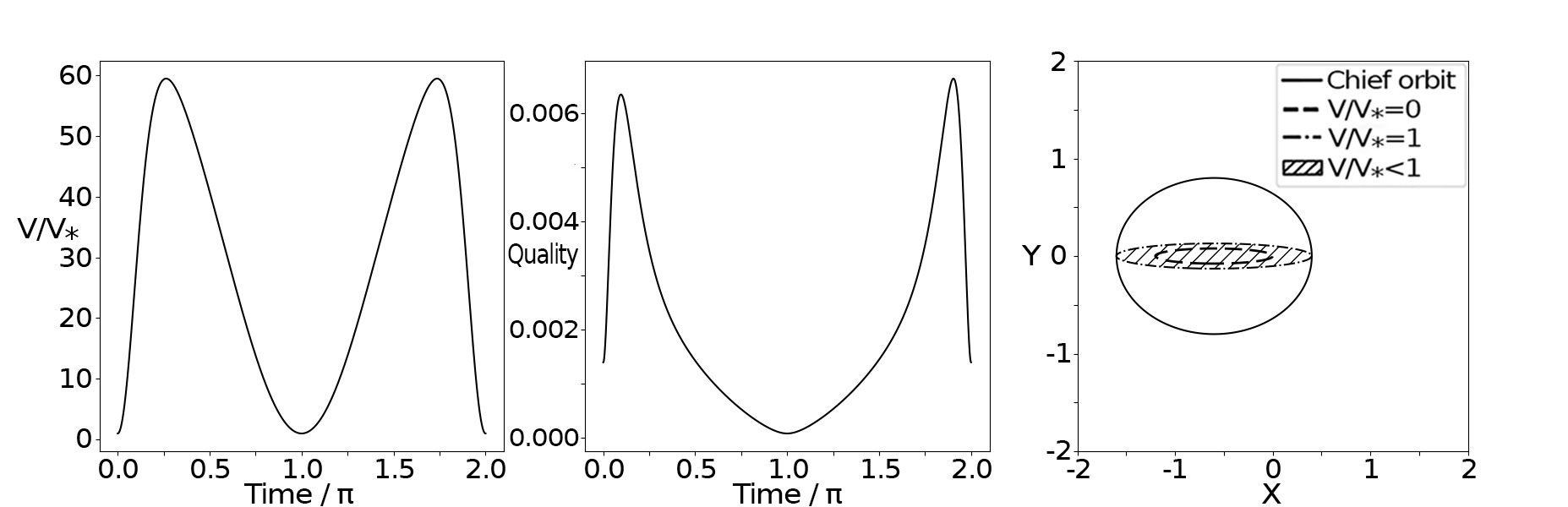}}
\caption*{\textbf{Fig.9. } An example of s significantly changing, not zeroed volume. Data see in the text.}
\end{figure}

\noindent

\begin{center}
\textit{3.2. An Example of a Detailed Analysis of the Evolution of a Formation}
\end{center}

Below an example of a detailed analysis of the evolution of the shape of a tetrahedron, starting from perihelion will be presented. Let the base of the tetrahedron (the triangle formed by the deputy spacecraft with numbers $m=1, 2, 3$) to be in the plane $x=x_{0} $ at perihelion. Moreover $x_{0} >0$ is large enough compared to the initial dimensions of the base, that the base always remains on one side of the chief with coordinates (0, 0, 0).  Let $u_{z0m} =0$, $w_{0m} =0$. In this case $S_{xx} x_{0m} =S_{xx} x_0$, $S_{yx} x_{0m} =S_{yx} x_0$ are independent of $m$,  $S_{xw} =S_{yw} =0$. The eccentricity is set to 0.6.

It can be seen from \eqref{GrindEQ__36_}, \eqref{GrindEQ__37_} that the plane of the base is always parallel to the \textit{z} axis. The angle \textit{$\varphi $} between the plane of the base and the \textit{y} axis will be determined by the formula ${\rm tg}\varphi =S_{xy} /S_{yy} $. Accordingly, this angle firstly increases when moving from perihelion, then returns to the initial value $\varphi =0$ at aphelion.  On the way from aphelion to perihelion, rotation occurs in the opposite direction.

The displacement of the base is divided into 2 types.

The first type is a general shift of the base without its deformation. On the way from perihelion to aphelion, the distance along $x$ direction from the top to the base increases up to $\approx 2$ times. On this path, the base also moves in $y$ direction towards negative $y$, and then towards positive one. At the aphelion, this shift is equal to 0 in accordance with the properties of $S_{yx} $. If $x_0$ substantially exceeds $|y_0|$ the discussed shift along $y$ can exceeds the size of the base.

The second type of displacement of the base is associated with $S_{xy}$, $S_{yx}$ and movement along $z$. It results in deformation of the base. The deformation consists of a slight compression along $y$ near the perihelion, and then a continuous extension along $y$ direction up to 6 times at the aphelion in accordance with the behavior of  $S_{yy}$. Under conditions $u_{z0m} =0$, motion along z-axis is described as follows: the coordinates $z_m$ decrease in absolute value to 0 at $X = 0$, then change sign and grow, reaching a maximum at the aphelion. This maximum is $R_{a} /R_p =(1+e)/(1-e)$ times greater than the values $z_m$ at the perihelion. At $X=0$ all 4 spacecraft are in the same plane $z = 0$.

\begin{center} \textbf{4. Discussion} \end{center}

In our opinion, the above indicates many advantages of the proposed mathematical apparatus for studying the tetrahedral spacecraft formation. The advantages are as follows.

\begin{enumerate}
\item  The obtained expressions are simpler compared to the Tschauner-Hempel formulae, requiring numerical analysis. Often, numerical integration of the equations of motion is simpler than using these formulas. In some cases, the approach proposed in this paper allows creating a qualitative, and sometimes even complete quantitative analysis of the spacecraft formation evolution without a computer.

\item  The use of orbital elements in the Tschauner-Hempel formulas is probably natural for describing the motion of celestial bodies for an observer located on Earth. In our case, the usual Cartesian coordinates that determine the relative positions of spacecraft are more natural.

\item  The proposed coordinate system is inertial. Its use does not result in the emergence of forces that are not associated with real physical fields. This can simplify the analysis of measurement results when studying the gravitational field of the Sun.

\item  The obtained fundamental solutions of the linearized equations of motion have a transparent physical meaning. They correspond to a certain type of deformation (rotations, time shifts, changes in eccentricity and semi-major axis), which makes it easier to understand the influence of various parameters.

\item  The condition of equality of the periods of revolution of spacecraft around the Sun, that is important for planning various missions, has a simple and clear form in the proposed approach.

\item  The obtained expression in the form of a second-degree polynomial for a quantity critical to mission planning -- the volume of a tetrahedron -- makes it possible to quickly determine critical points at which the volume vanishes or reaches extremes without resorting to numerical calculations with a fine time step.

\item  For various mission planning purposes, it may be reasonable to firstly determine the coefficients $c_{1-5} $ ensuring the desired volume behavior. This can significantly narrow the search area for the optimal spacecraft coordinates for a given mission.

\item  Writing the expression for the volume of a tetrahedron as mixed products in the space of spacecraft numbers has obvious symmetries, which can be used to construct the desired spacecraft formation.

\item  The proposed approach can be extended, using standard perturbation theory, to take nonlinear terms and non-gravitational forces into account.

\item  When planning various missions, the proposed approach can be used to roughly determine the region of interest and then search for the best parameters options within that region, using more accurate, albeit computationally-intensive, methods.
\end{enumerate}

Let us discuss the applicability of the linear approximation. If the characteristic distance between spacecraft is $r_0 =1000$ km, and the orbit size is $R_0 =1.5\cdot 10^{8} $ km, then the expansion parameter $r_0/R_0$ is of the order  of $10^{-5} $. The 2nd order correction to the distance between spacecraft $R_{0} (r_{0} /R_{0} )^{2} $ is of the order of tens of meters. The characteristic relative velocity $V_{0} (r_{0} /R_{0} )$ is of the order of tens of centimeters per second, i.e. small.  Here $V_0 =30$ km/sec is the characteristic velocity of the chief. Nonlinear corrections to the velocity $V_0 (r_0/R_0)^2$ are negligibly small. The given estimates are valid for moderate eccentricities. The ratio of the minor semiaxis, or the distance from the spacecraft to the Sun at perihelion, acts as a small parameter and worsens the estimates of the applicability of the linear approximation. At $e$ close to 1, the ratio $r_0$ to the minor semi-axis or to the distance from the spacecraft to the Sun at perihelion will act as a small parameter and it worsens the estimates of the applicability of the linear approximation. However, such orbits are of a little interest.

Note that the influence of nonlinear terms, as well as non-gravitational forces or the attraction of other planets, can be taken into account using standard perturbation theory. The last two corrections must considered for the motion of the chief as well.

\begin{center} \textbf{5. Conclusion} \end{center}

Thus, new effective tool for mathematical description of a 4-spacecraft formation is proposed in this paper. A significant accomplishment of this study is the formula obtained for the volume of a tetrahedron expressed as a polynomial in the Cartesian coordinates of the chief.  In the case of equal orbital periods, this polynomial is of the second degree, indicating that the volume can vanish from zero to four times per period.

Further development of this work could include determining the optimal formation for a specific mission, such as a mission to test gravity theories.  This requires formulating the specific requirements for the spacecraft relative positions necessary for making measurements. This is a complex task.  Studying the various properties of the resulting metamathematical object -- a tetrahedron that transforms according to a linear approximation -- may be interesting and useful.  Space research development may impose various requirements on this formation.  Deriving a formula for the quality of a tetrahedron similar to the formula for its volume would also be useful.

\noindent {}

\noindent \textbf{CRediT authorship contribution statement}

\noindent \textbf{Vladimir Zhukov:} Conceptualization (lead), Methodology (lead), Writing -- original draft (lead), Writing -- review \& editing (equal). \textbf{Nikolai Iakovlev:} Conceptualization (supporting), Methodology (supporting), Visualization (equal), Writing -- review \& editing (equal), Project administration (lead). \textbf{Alexander Bochkarev:} Conceptualization (supporting), Methodology (supporting), Visualization (equal), Writing -- review \& editing (equal). \textbf{Nikita Logvinenko:} Conceptualization (supporting), Methodology (supporting), Software (lead), Writing -- review \& editing (equal). \textbf{Sergei Kurchev:} Conceptualization (supporting), Methodology (supporting), Software (supporting). \textbf{Vlas Karavaikin:} Conceptualization (supporting). \textbf{Ivan Radko:} Conceptualization (supporting).

\noindent \textbf{Funding}

\noindent This research received no external funding.

\noindent \textbf{Declaration of competing interest}
\noindent The authors declare that they have no known competing financial interests or personal relationships that could have appeared to influence the work reported in this paper.

\noindent {}

\begin{center} \textbf{Appendix A. Nomenclature} \end{center}

\noindent $c_1 ,c_2 ,c_3 ,c_4,c_5$ = coefficients in the polynomial for the tetrahedron volume

\noindent $D$ =  determinant for the tetrahedron volume

\noindent \textit{e} = eccentricity of the reference orbit

\noindent $G=\gamma M$ = the gravity multiplier

\noindent $M$ = the mass of the Sun

\noindent $m$ = the number of the deputy spacecraft, \textit{m}=1,2,3

\noindent $Q$ = quality of the tetrahedron

\noindent ${\bf R}$ = vector of the chief position

\noindent $R$ = the distance between the chief and the Sun

\noindent ${\bf R}_m$= vectors of positions of the deputies

\noindent $R_p$ = the radius of the reference orbit at perihelion

\noindent ${\bf r}_m$ = vectors of positions of the deputy relatively to the position of the chief, ${\bf r}_m={\bf R}_m -{\bf R}$

\noindent $t$ = time

\noindent ${\bf u}_m$ = relative velocities vectors of the $m$-th deputy, ${\bf u}_m =\dot{{\bf r}}_m$

\noindent $u_{xm} ,u_{ym} ,u_{zm} $ = Cartesian components of relative ${\bf u}_m$

\noindent ${\bf V}$ = the chief velocity, ${\bf V}=\dot{{\bf R}}$

\noindent $V$ = magnitude of ${\bf V}$

\noindent V = volume of tetrahedron

\noindent $V_p$  = the speed of the chief at perihelion

\noindent $w_m$ = a value, characterizing the $m$-th deputy relative velocity component perpendicular to the velocity of the chief

\noindent $X,Y,Z$ = Cartesian coordinates of the chief, ${\bf R}=[X,Y,Z]$, $Z=0$

\noindent $x_m ,y_m ,z_m$ = relative Cartesian coordinates of the $m$-th deputy, ${\bf r}_m =[x_m, y_m,z_m]$

\noindent ${\bf x}$, ${\bf y}$, ${\bf z}$, ${\bf w}$  = vectors in the space of deputies numbers with components ${\bf x}=[x_1 ,x_2 ,x_3]$, ${\bf y}=[y_1 ,y_2 ,y_3]$, \dots

\noindent ${\bf x}_0$, ${\bf y}_0$, ${\bf z}_0$, ${\bf u}_{z0}$, ${\bf w}_0$ = values of vectors ${\bf x}$, ${\bf y}$, ${\bf z}$, ${\bf u}_z$, ${\bf w}$ at perihelion

\noindent $\alpha _m$, $\beta _m$, $\chi _m$, $\tau _m$, $\eta _m$, $\upsilon _m$ = constants of motions of the $m$-th deputy

\noindent $\pmb \alpha$, $\pmb \beta$, \dots 
 = vectors in the space of deputies numbers with components $\pmb \alpha =[\alpha _1 ,\alpha _2 ,\alpha _3]$, $\pmb \beta  =[\beta _1 ,\beta _2 ,\beta _3 ]$,\dots

\noindent $\gamma $ = the gravity constant

\noindent {}

\begin{center} \textbf{Appendix B. Calculation $\partial X/\partial e$, $\partial Y/\partial e$} \end{center}

\noindent These derivatives can be calculated as follows. Eq.-s (4), (5) give
$$\frac{\partial X}{\partial e} =-\left(1+\frac{\partial \xi }{\partial e} \sin \xi \right),\ \ \frac{\partial Y}{\partial e} =\sqrt{1-e^{2} } \cos \xi \frac{\partial \xi }{\partial e} -\frac{e}{\sqrt{1-e^{2} } } \sin \xi $$
$$\frac{\partial \xi }{\partial e} -\sin \xi -e\cos \xi \frac{\partial \xi }{\partial e} =0$$
Therefore, using also (4)
$$\left. \frac{\partial \xi }{\partial e} \right|_{t={\rm const}} =\frac{\sin \xi }{1-e\cos \xi } =\frac{\sin \xi }{R} $$
Respectively, using Eq. (4), (6)
$$\frac{\partial X}{\partial e} =-\left(1+\frac{\sin ^2 \xi }{R} \right)=-\left(1+\frac{Y^{2} }{R(1-e^2)} \right),$$
$$\frac{\partial Y}{\partial e} =\sqrt{1-e^{2} } \cos \xi \frac{\sin \xi }{R} -\frac{e}{\sqrt{1-e^{2} } } \sin \xi =\frac{XY}{R(1-e^{2} )}$$

\noindent {}

\begin{center} \textbf{Appendix C. Calculation $\partial X/\partial a$, $\partial Y/\partial a$} \end{center}

The dimensional solutions corresponding to the elliptical orbits with semi-major axis $a$ are
\begin{equation*}
X=a(\cos \xi -e),\ \ Y=a\sqrt{1-e^{2} } \sin \xi , \eqno {(A1)}
\end{equation*}
\begin{equation*}
\xi -e\sin \xi =t/a^{3/2} \eqno {(A2)}
\end{equation*}

\noindent Eq. (A1) gives
$$\frac{\partial X}{\partial a} =\frac{\partial (a(\cos \xi -e))}{\partial a} =\frac{X}{a} +\frac{\partial X}{\partial \xi } \frac{\partial \xi }{\partial a} =\frac{X}{a} +\dot{X}\left(\frac{\partial \xi }{\partial t} \right)^{-1} \frac{\partial \xi }{\partial a} $$
From Eq. (A2) follows
$$\frac{\partial \xi }{\partial t} -e\cos \xi \frac{\partial \xi }{\partial t} =a^{-3/2},\ \  \frac{\partial \xi }{\partial a} -e\cos \xi \frac{\partial \xi }{\partial a} =-\frac{3}{2} \frac{t}{a^{5/2} } $$
These give
$$\left(\frac{\partial \xi }{\partial t} \right)^{-1} \frac{\partial \xi }{\partial a} =-\frac{3}{2} \frac{t}{a} $$
Thus $$\frac{\partial X}{\partial a} =\frac{X}{a} -\frac{3}{2} \frac{\dot{X}}{a} t$$
In the dimensionless case $a=1$, this gives
\begin{center}
$\partial X/\partial a=X-(3/2)t\dot{X}$ and analogously $\partial Y/\partial a=Y-(3/2)t\dot{Y}$.\end{center}

\noindent {}

\begin{center} \textbf{Appendix D. The Proof of the Equality $\frac{XY\dot{X}+Y^{2} \dot{Y}}{R(1-e^{2} )} +\dot{Y}=\frac{2e+X}{\sqrt{1-e^2 } } $} \end{center}

\noindent The left side $L$ of the equality under discussion can be rewritten using Eq.-s (12), (9)
$$L=\frac{XY\dot{X}+Y^{2} \dot{Y}}{R(1-e^{2} )} +\dot{Y}=\frac{(X\dot{X}+Y\dot{Y})Y}{R(1-e^{2} )} +\dot{Y}
=\frac{eY^{2} }{R(1-e^{2} )^{3/2} } +\frac{(X+e)}{R} \sqrt{1-e^{2} } $$
Next, using Eq. (13),
\[L=\frac{e\left(1-(X+e)^{2} \right)}{R\sqrt{1-e^{2} } } +\frac{(X+e)\sqrt{1-e^{2} } }{R} =\frac{e\left(1-(X+e)^{2} \right)+(X+e)(1-e^{2} )}{R\sqrt{1-e^{2} } } .\]
Using Eq. (6)
$$L=\frac{e\left(1-\left(\frac{1-R}{e} \right)^{2} \right)+\frac{1-R}{e} (1-e^{2} )}{R\sqrt{1-e^2} } =\frac{1-R+e^{2} }{e\sqrt{1-e^{2} } } .$$
Going back from $R$ to $X$ using Eq. (6) the desired $L=(2e+X)/\sqrt{1-e^2}$ is obtained.

\noindent {}

\begin{center} \textbf{Appendix E. The Existence of Spacecraft Formation Given Any Values of the Coefficients $c_{1-5}$ in the Formula for Tetrahedron Volume} \end{center}

Let us show that for any non-zero and non-parallel $\alpha $ and $\beta $, i.e. $\pmb \alpha ^2 \pmb \beta ^2 -(\pmb \alpha \pmb \beta )^2 \ne 0$, and for any $c_1$, $c_2$, $c_3$, $c_4$, $c_5$ (33), including
$c_1 =c_2 =c_3 =c_4 =c_5=0$, exist $\pmb \tau $, $\pmb \eta $, $\pmb \chi $, not all equal to zero  ($\pmb \tau ^2 +\pmb \eta ^2 +\pmb \chi ^2 \ne 0$) which give these $c_1$, $c_2$, $c_3$, $c_4$, $c_5$.

To do this, the vectors ${\bf A}_0$, ${\bf B}_0$, ${\bf C}_0$,  lying in the plane $\pmb \alpha $-$\pmb \beta $, and the scalars $A_\perp $, $B_\perp $, $C_\perp $  are used according to the formulae
\begin{equation*}
e[\pmb \tau \times \pmb \chi ]={\bf A}_0 +A_{\bot } [\pmb \alpha \times \pmb \beta ],\ \ [\pmb \tau \times \pmb \eta ]={\bf B}_0 +B_{\bot } [\pmb \alpha \times \pmb \beta ],\ \ \sqrt{1-e^{2} } [\pmb \chi \times \pmb \eta ]={\bf C}_0 +C_{\bot } [\pmb \alpha \times \pmb \beta ] \eqno {(A3)}
\end{equation*}

Substitution (A3) into
\begin{equation*}
{\rm V}=|D|/6,\ D=\left(\pmb \alpha X+\pmb \beta Y\right)\left([\pmb \chi \times \pmb \eta ]X+[\pmb \tau \times \pmb \chi ]\frac{eY}{\sqrt{1-e^{2} } } +[\pmb \tau \times \pmb \eta ]\frac{2e+X}{\sqrt{1-e^{2} } } \right) \eqno {(A4)}
\end{equation*}
 gives ${\rm V}=(6\sqrt{1-e^2})^{-1} \left(\pmb \alpha X+\pmb \beta Y\right)\left(\left({\bf C}_0 +{\bf B}_0 \right)X+{\bf A}_0 Y+2e{\bf B}_0 \right)$. Removing brackets and comparing this with (33), one concludes that it is necessary to satisfy equalities
\begin{equation*}
(\pmb \alpha {\bf C}_0 )+(\pmb \alpha {\bf B}_0)=6c_1 \sqrt{1-e^2}
\eqno {(A5)}
\end{equation*}
\begin{equation*}
(\pmb \beta {\bf A}_0 )=6c_2 \sqrt{1-e^2}
\eqno {(A6)}
\end{equation*}
\begin{equation*}
(\pmb \alpha {\bf A}_0)+(\pmb \beta {\bf B}_0 )+(\pmb \beta {\bf C}_0)=6c_3 \sqrt{1-e^2}
\eqno {(A7)}
\end{equation*}
\begin{equation*}
(\pmb \alpha {\bf B}_0)=\frac{3\sqrt{1-e^{2} } }{e} c_4
\eqno {(A8)}
\end{equation*}
\begin{equation*}
(\pmb \beta {\bf B}_0)=\frac{3\sqrt{1-e^{2} } }{e} c_5
\eqno {(A9)}
\end{equation*}

The values $A_{\bot }$, $B_{\bot }$, $C_{\bot }$ can be arbitrary.

It is easy to get from Eq.-s (A5), (A7), (A8)
\begin{equation*}
(\pmb \alpha {\bf C}_0)=\left(6c_1 -(3/e)c_4 \right)\sqrt{1-e^2}
\eqno {(A10)}
\end{equation*}
\begin{equation*}
(\pmb \alpha {\bf A}_0)+(\pmb \beta {\bf C}_0)=(6c_3 -(3/e)c_5)\sqrt{1-e^2}
\eqno {(A11)}
\end{equation*}

Thus, specifying arbitrary $(\pmb \alpha {\bf A}_0)$ or $(\pmb \beta {\bf C}_0)$, projections of  vectors ${\bf A}_0$, ${\bf B}_0$, ${\bf C}_0$ on the vectors  $\pmb \alpha $ and $\pmb \beta $ can be found.  Respectively, the vectors themselves can be found. Note, that $(\pmb \alpha {\bf A}_0)$ and $(\pmb \beta {\bf C}_0)$ can be nonzero even if all $c_{n} =0$.

Below an example of building of $\pmb \tau $, $\pmb \chi $, $\pmb \eta $ using given ${\bf A}_0$, ${\bf B}_0$, ${\bf C}_0$ will be presented. To do this, we introduce components $\pmb \tau _0$, $\pmb \chi _0$, $\pmb \eta _0$ of vectors $\pmb \tau $, $\pmb \chi $, $\pmb \eta $, lying in the plane $\pmb \alpha $-$\pmb \beta $ and present them in the form
$$\tau =\pmb \tau _0 +\tau _{\bot } [\pmb \alpha \times \pmb \beta ], \ \ \pmb \chi =
\pmb \chi _0 +\chi _{\bot } [\pmb \alpha \times \pmb \beta ], \ \ \pmb \eta =\pmb \eta _0
 +\eta _{\bot } [\pmb \alpha \times \pmb \beta ]$$

Let us consider a special case $\pmb \tau _0 =0$, $\tau _{\bot } \ne 0$. From the first and second expressions of (A3) can be found
$$\pmb \chi _0 =-\frac{[\pmb \tau \times {\bf A}_0]}{e\tau ^2},
\ \pmb \eta _0 =-\frac{[\pmb \tau \times {\bf B}_0 ]}{\tau ^2} $$
From the last expression of (A3) can be found
$$[(-[\pmb \tau \times {\bf A}_0](e\tau ^{2} )^{-1} +\chi _{\bot } [\pmb \alpha \times \pmb \beta ])
\times (-[\pmb \tau \times {\bf B}_0]\tau ^{-2} +\eta _{\bot } [\pmb \alpha \times \pmb \beta ])]
={\bf C}_0+C_{\bot } [\pmb \alpha \times \pmb \beta ]$$
next, taking into account $\pmb \tau =\tau _{\bot } [\pmb \alpha \times \pmb \beta ]\bot {\bf A}_0 ,{\bf B}_0 ,{\bf C}_0$,
\begin{equation*}
(e\tau _{\bot } )^{-1} ({\bf A}_0 \eta _{\bot } -{\bf B}_0 e\tau _{\bot } \chi _{\bot } )-\frac{\pmb \tau (\pmb \tau [{\bf A}_0 \times {\bf B}_0])}{e\tau ^4}
={\bf C}_0 +C_{\bot } \frac{\pmb \tau }{\tau _{\bot } }
\eqno {(A12)}
\end{equation*}

Hence
$$C_{\bot } =-\frac{\tau _{\bot } (\pmb \tau [{\bf A}_0 \times {\bf B}_0])}{e\tau ^4} $$
Multiplying (A12) by $\pmb \alpha $ and $\pmb \beta $, and taking into account (A11), we obtain
\begin{equation*}
(\pmb \alpha {\bf A}_0)\eta _{\bot } -(\pmb \alpha {\bf B}_0 )e\tau _{\bot } \chi _{\bot } =(\pmb \alpha {\bf C}_0)e\tau _{\bot } \equiv e\tau _{\bot } \left(6c_1 -(3/e)c_4 \right)\sqrt{1-e^2}
\eqno {(A13)}
\end{equation*}
\begin{equation*}
(\pmb \beta {\bf A}_0)\eta _{\bot } -(\pmb \beta {\bf B}_0)e\tau _{\bot } \chi _{\bot } =e\tau _{\bot } \left(-(\pmb \alpha {\bf A}_0)+(6c_3 -(3/e)c_5 )\sqrt{1-e^{2} } \right)
\eqno {(A14)}
\end{equation*}

The parameter $(\pmb \alpha {\bf A}_0)$ is arbitrary in these equations for $\eta _{\bot }$, $\chi _{\bot } $. It can be chosen so that the equations are solvable. Let us consider various situations.

\begin{enumerate}
\item  $(\pmb \alpha {\bf A}_0)=(6c_{3} -(3/e)c_{5} )\sqrt{1-e^{2} }$, $\chi _{\bot } =\left((\pmb \alpha {\bf A}_0)\eta _{\bot } -(\pmb \alpha {\bf C}_0)e\tau _{\bot } \right)/\left((\pmb \alpha {\bf B}_0)e\tau _{\bot } \right)$ and  arbitrary $\eta _{\bot }$, $\tau _{\bot } \ne 0$  can be a solution of (A13), (A14) in the case $(\pmb \beta {\bf A}_0)=(\pmb \beta {\bf B}_0)=0$, $(\pmb \alpha {\bf B}_0)\ne 0$.

\item  In the case$(\pmb \beta {\bf B}_0)\ne 0$ the determinant for the linear equations (A13), (A14) for $\eta _{\bot } $ and $\chi _{\bot } $ can be made non-zero by not unique choice of $(\pmb \alpha {\bf A}_0)$. Respectively, the equations can be solved.

\item  In the case $(\pmb \beta {\bf B}_0)=0$, $(\pmb \alpha {\bf B}_0)=0$ $\Rightarrow {\bf B}_0 =0$ the solution with $\pmb \tau _0=0$, $\tau _{\bot } \ne 0$ might require a different approach, as shown below. In this case formulae (A5)--(A11) have a view
\begin{equation*}
(\pmb \alpha {\bf C}_0)=6c_1 \sqrt{1-e^2}
\eqno {(A15)}
\end{equation*}
\begin{equation*}
(\pmb \beta {\bf A}_0)=6c_2 \sqrt{1-e^2}
\eqno {(A16)}
\end{equation*}
\begin{equation*}
(\pmb \alpha {\bf A}_0)+(\pmb \beta {\bf C}_0)=6c_3 \sqrt{1-e^2}
\eqno {(A17)}
\end{equation*}
\end{enumerate}

Taking arbitrary $(\pmb \alpha {\bf A}_0)$ and using Eq.-s (A15)--(A17) the vectors ${\bf A}_0$, ${\bf C}_0$ can be found for given $c_1$, $c_2$, $c_3$. It is easy to see that a class of solutions
$$\pmb \chi =\chi _{\bot } [\pmb \alpha \times \pmb \beta ],\ \pmb \tau =\frac{[\pmb \chi \times {\bf A}_0]}{e\chi ^2 }, \  \pmb \eta =-\frac{[\pmb \chi \times {\bf C}_0]}{\chi ^{2} \sqrt{1-e^{2} } } $$
satisfies (A15)--(A17) and (A3) in which ${\bf B}_0=0$. In that case $B_{\bot } [\pmb \alpha \times \pmb \beta ]=[\pmb \tau \times \pmb \eta ]$, $A_{\bot } =0$, $C_{\bot } =0$.

This result shows, that the formation in which all 4 spacecraft permanently belong to a single plane, but their orbits don't lie in one plane is possible.

\noindent {}

\noindent \textbf{References}

\noindent [1] T. Karlsson, G.T. Marklund, S. Figueiredo, T. Johansson, S. Buchert, Separating spatial and temporal variations in auroral electric and magnetic fields by Cluster multipoint measurements, \textit{Annales Geophysicae}, 22(7) (Jul. 2004) 2463--2472. https://doi.org/10.5194/angeo-22-2463-2004

\noindent [2] S. Shestakov, M. Ovchinnikov, Y. Mashtakov, Analytical approach to construction of tetrahedral satellite formation, \textit{Journal of Guidance, Control, and Dynamics}, 42(12) (Dec. 2019) 2600--2614.

\noindent https://doi.org/10.2514/1.G003913

\noindent [3] M. Ovchinnikov, Y. Mashtakov, S. Shestakov, Lyapunov-Based Control via Atmospheric Drag for Tetrahedral Satellite Formation, \textit{Mathematics}, 12(2) (Jan. 2024) 189. https://doi.org/10.3390/math12020189

\noindent [4] C.P. Escoubet, M. Fehringer, M. Goldstein, Introduction the cluster mission, \textit{Annales Geophysicae}, 19(10/12) (Sep. 2001)  1197--1200. https://doi.org/10.5194/angeo-19-1197-2001

\noindent [5] C.P. Escoubetand, R. Schmidt, Cluster II: Plasma measurements in three dimensions, \textit{Advances in Space Research}, 25(7-8) (Jan. 2000) 1305--1314. https://doi.org/10.1016/S0273-1177(99)00639-0

\noindent [6] D. Southwood, W.H. Stanley, S.W. Cowley, S. Mitton, Eds. \textit{Magnetospheric Plasma Physics: The Impact of Jim Dungey's Research}, 41 (Aug. 2015), Springer. DOI 10.1007/978-3-319-18359-6

\noindent [7] G. Paschmann, C.P. Escoubet, S.J. Schwartz, S. Haaland, Eds. \textit{Outer magnetospheric boundaries: Cluster results}, 20 (Dec. 2005), Springer Science \& Business Media.

\noindent [8] C.P. Escoubet, R. Schmidt, M.L. Goldstein, Cluster--science and mission overview, \textit{Space Science Reviews}, 79(1) (Jan. 1997) 11--32. https://doi.org/10.1023/A:1004923124586

\noindent [9] A. Balogh, M.W. Dunlop, S.W.H. Cowley, D.J. Southwood, J.G. Thomlinson, K.H. Glassmeier, G. Musmann, H. Luhr, S. Buchert, M.H. Acuna, D.H. Fairfield, The Cluster magnetic field investigation, \textit{Space Science Reviews}, 79 (Jan. 1997) 65--91. https://doi.org/10.1023/A:1004970907748

\noindent [10] M.G. Taylor, C.P. Escoubet, H. Laakso, A. Masson, M. Hapgood, T. Dimbylow, J. Volpp, S. Sangiorgi, M.L. Goldstein, The Science of the Cluster Mission, \textit{Magnetospheric Plasma Physics: The Impact of Jim Dungey's Research}, (2015) 159--179, Springer International Publishing. https://doi.org/10.1007/978-3-319-18359-6\_8

\noindent [11] B.H. Mauk, R.W. McEntire, R.A. Heelis, R.F., Pfaff Jr, Magnetospheric multiscale and global electrodynamics missions, \textit{Sun-Earth Plasma Connections}, 109 (Jan. 1999) 225--235. https://doi.org/10.1029/GM109

\noindent [12] J.L. Burch, T.E. Moore, R.B. Torbert, B.H. Giles, Magnetospheric multiscale overview and science objectives, \textit{Space Science Reviews}, 199 (Mar. 2016) 5--21. https://doi.org/10.1007/s11214-015-0164-9

\noindent [13] S.G. Turyshev, S.W. Chiow, N. Yu, Searching for new physics in the Solar System with tetrahedral spacecraft formations, \textit{Physical Review D}, 109(8) (Apr. 2024) 084059. https://doi.org/10.1103/PhysRevD.109.084059

\noindent [14] N. Yu, S.-w. Chiow, J. Gleyzes, P. Bull, O. Dore J. Rhodes, J. Jewell, E. Huff, H. Muller, \textit{Direct probe of dark energy interactions with a Solar System laboratory, The Final Report NIAC Phase I proposal, }2018.

\noindent  URL: https://ntrs.nasa.gov/api/citations/20190002500/downloads/20190002500.pdf

\noindent [15] D. Benisty, Testing modified gravity via Yukawa potential in two body problem: Analytical solution and observational constraints, \textit{Physical Review D}, 106(4) (Aug. 2022) 043001.

\noindent https://doi.org/10.1103/PhysRevD.106.043001

\noindent [16] A. Padilla, P.M. Saffin, S.Y. Zhou, Bi-galileon theory I: Motivation and formulation, \textit{Journal of High Energy Physics}, 2010(31) (Dec. 2010) 1--26. https://doi.org/10.1007/JHEP12(2010)031

\noindent [17] M. Trodden, K. Hinterbichler, Generalizing galileons, \textit{Classical and Quantum Gravity}, 28(20) (Oct. 2011) 204003. https://doi.org/10.1088/0264-9381/28/20/204003

\noindent [18] C. Deffayet, D.A. Steer, A formal introduction to Horndeski and Galileon theories and their generalizations, \textit{Classical and Quantum Gravity}, 30(21) (Oct. 2013) 214006. https://doi.org/10.1088/0264-9381/30/21/214006

\noindent [19] N. Chow, J. Khoury, Galileon cosmology, \textit{Physical Review D---Particles, Fields, Gravitation, and Cosmology}, 80(2) (Jul. 2009) 024037. https://doi.org/10.1103/PhysRevD.80.024037

\noindent [20] J. Tschauner, P. Hempel, Rendezvous zu einem in elliptischer Bahn umlaufenden Ziel, \textit{Astronautica Acta}, 11(2) (Jan. 1965) 104--109.

\noindent [21] C. Lane, P. Axelrad, Formation design in eccentric orbits using linearized equations of relative motion, \textit{Journal of Guidance, Control, and Dynamics}, 29(1) (Jan. 2006) 146--160. https://doi.org/10.2514/1.13173

\noindent [22] M. Bando, A. Ichikawa, Graphical generation of periodic orbits of Tschauner-Hempel equations, \textit{Journal of guidance, control, and dynamics}, 35(3) (May 2012) 1002--1007. https://doi.org/10.2514/1.56326

\noindent [23] P. Sengupta, S.R. Vadali, Relative motion and the geometry of formations in Keplerian elliptic orbits with arbitrary eccentricity, \textit{Journal of guidance, control, and dynamics}, 30(4) (Jul. 2007) 953--964.

\noindent https://doi.org/10.2514/1.25941

\noindent [24] K. Yamanaka, F. Ankersen, New state transition matrix for relative motion on an arbitrary elliptical orbit, \textit{Journal of guidance, control, and dynamics}, 25(1) (Jan. 2002) 60--66. https://doi.org/10.2514/2.4875

\noindent [25] L. D. Landau, E. M. Lifshitz, \textit{Mechanics }3rd ed. Butterworth-Heinemann, 1976.

\noindent [26] J. Guzman, C. Schiff, A preliminary study for a tetrahedron formation: quality factors and visualization, \textit{AIAA/AAS Astrodynamics Specialist Conference and Exhibit}, Aug. 2002, p. 4637.

\noindent https://doi.org/10.2514/6.2002-4637

\noindent [27] P.W. Daly, The tetrahedron quality factors of CSDS, \textit{Max Planck Inst. fur Aeronomie Tech. Rept. MPAe--W--100--94--27, Katlenburg-Lindau, D--37191, Germany}, 7 Jun. 1994.

\noindent [28] P. Robert, A. Roux, Accuracy of the estimate of J via multipoint measurements, \textit{Space Plasma Physics Investigation by Cluster and Regatta}, May 1990, pp. 29--35.

\noindent [29] P. Robert, A. Roux, O. Coeur-Joly, September. Validity of the estimate of the current density along Cluster orbit with simulated magnetic data, \textit{Proceedings of the Cluster Workshops, Data Analysis Tools (Braunschweig, Germany, 28-30 September 1994) and Physical Measurements and Mission-Oriented Theory (Toulouse, France, 16017 November 1994). European Space Agency. Edited by K.-H. Glassmeier, U. Motschmann, and R. Schmidt, 1995}, Vol. 371, Sep. 1995, p. 229.

\noindent [30] S.P. Hughes, Orbit design for phase I and II of the Magnetospheric Multiscale Mission, \textit{27th Annual Guidance and Control Conference}, NASA Goddard Space Flight Center; Greenbelt, MD, United States, 1 Jan.  2004. Vol. 118. P. 255--274.

\noindent [31] S. Hughes, Formation design and sensitivity analysis for the magnetospheric multiscale mission (mms), \textit{AIAA/AAS Astrodynamics Specialist Conference and Exhibit}, 2008, p. 7357. https://doi.org/10.2514/6.2008-7357

\noindent [32] V.N. Parthasarathy, C.M. Graichen, A.F. Hathaway, A comparison of tetrahedron quality measures, \textit{Finite Elements in Analysis and Design}, 15(3) (Jan. 1994) 255--261. https://doi.org/10.1016/0168-874X(94)90033-7

\noindent [33] A. Liu, B., Joe, Relationship between tetrahedron shape measures, \textit{BIT Numerical Mathematics}, 34(2) (Jun. 1994) 268--287. https://doi.org/10.1007/BF01955874

\noindent [34] A. Liu, B. Joe, On the shape of tetrahedra from bisection, \textit{Mathematics of computation}, 63(207) (1994) 141--154.  https://doi.org/10.1090/S0025-5718-1994-1240660-4

\noindent

\end{document}